\documentclass{aastex61}

\usepackage{amsmath}
\usepackage{multirow}

\newcommand{\msunyr} {M_\odot~{\rm yr}^{-1}}

\newcommand{\scl}	{\Sigma_{\rm cl}}
\newcommand{\gcm}	{{\rm g}\,{\rm cm}^{-2}}
\newcommand{\chisq}     {\chi^2}
\newcommand{\chisqmin}     {\chi^2_\mathrm{min}}
\newcommand{\mc}        {M_c}
\newcommand{\ms}        {m_*}
\newcommand{\inc}       {\theta_\mathrm{view}}
\newcommand{\muinc}  {\mu_\mathrm{view}}
\newcommand{\av}        {A_V}
\newcommand{\fnu}       {F_\nu}
\newcommand{\fnuobs}    {F_{\nu,\mathrm{obs}}}
\newcommand{\fnumod}    {F_{\nu,\mathrm{mod}}}
\newcommand{\fnumodav}    {F_{\nu,\mathrm{mod},\mathrm{ext}}}
\newcommand{\fnufit}    {F_{\nu,\mathrm{fit}}}
\newcommand{\sigmau}     {\sigma_u}
\newcommand{\sigmal}     {\sigma_l}

\newcommand{\thetaw} {\theta_{w,\mathrm{esc}}}
\newcommand{\menv} {M_\mathrm{env}}
\newcommand{\lbol} {L_\mathrm{bol}}
\newcommand{\tbol} {T_\mathrm{bol}}
\newcommand{\linc} {L_\mathrm{inc}}
\newcommand{\ltot} {L_\mathrm{tot}}

\shorttitle{Massive Star Formation SED Models}
\shortauthors{Zhang \& Tan}

\begin{document}

\title{Radiation Transfer of Models of Massive Star Formation. IV. The Model Grid and Spectral Energy Distribution Fitting}

\author{Yichen Zhang}
\affil{The Institute of Physical and Chemical Research (RIKEN), Hirosawa 2-1, Wako-shi, Saitama, 351-0198, Japan;\\ yczhang.astro@gmail.com}
\author{Jonathan C. Tan}
\affil{Departments of Astronomy \& Physics, University of Florida, Gainesville, FL 32611, USA;\\ jctan.astro@gmail.com}

\begin{abstract}
We present a continuum radiative transfer model grid for fitting
observed spectral energy distributions (SEDs) of massive protostars.
The model grid is based on the paradigm of core accretion theory for
massive star formation with pre-assembled gravitationally-bound cores
as initial conditions. In particular, following the Turbulent
Core Model, initial core properties are set primarily
by their mass and the pressure of their ambient clump.
We then model the evolution of the protostar and its surround structures
in a self-consistent way. 
The model grid contains about 9000 SEDs with 4 free parameters:
initial core mass, the mean surface density of the environment, 
the protostellar mass, and the inclination. 
The model grid is used to fit observed SEDs via
$\chisq$ minimization, with the foreground extinction additionally estimated.
We demonstrate the fitting process and
results using the example of massive protostar G35.20-0.74.
Compared with other SED model grids currently used for massive star
formation studies, in our model grid, the properties of
the protostar and its surrounding structures are more physically
connected, which reduces the dimensionality of the parameter spaces
and the total number of models. This excludes possible fitting of
models that are physically unrealistic or that are not internally
self-consistent in the context of the Turbulent Core Model. Thus, this
model grid serves not only as a fitting tool to estimate properties of
massive protostars, but also as a test of core accretion theory.  
The SED model grid is publicly released with this paper.
\end{abstract}


\section{Introduction}
\label{sec:intro}

Massive stars impact many areas of astrophysics, yet there is still no
consensus on how they form.  Theories range from Core Accretion
models, i.e., scaled-up versions of low-mass star formation (e.g., the
Turbulent Core Model of \citealt[]{MT02, MT03}), to Competitive
Accretion models at the crowded centers of forming star clusters
(e.g., \citealt[]{Bonnell01}; \citealt[]{Wang10}), to Protostellar
Collisions (\citealt[]{Bonnell98}; \citealt[]{Bally05}). Such
confusion is partly due to the observational difficulties caused by
the relative rarity and the typically large distances ($\gtrsim 1$ kpc) of massive protostars,
highly crowded environments, and high extinctions.  The environments
of massive star formation are observed to be massive, dense gas clumps
with mass surface densities of $\scl\approx 1~\gcm$, which corresponds
to a visual extinction of about $\av\approx 200$ mag (e.g., see
\citealt[]{Tan14} for a review).

Analysis of broad-band spectral energy distributions (SED) composed of
total fluxes from NIR to FIR/sub-mm of massive protostars, via
radiative transfer (RT) modeling, is a primary way to understand the
properties of massive protostars, being efficient for large samples. A
number of RT models have been developed to compare with observations.
\citet[]{Robitaille06,Robitaille07} developed a large model grid to
fit the SEDs of young stellar objects (YSOs).  However, their model
grid was mainly developed for low-mass star formation, without
coverage of the parameter space needed for massive star formation,
such as very high accretion rates resulting from high mass surface
density environments.  A similar RT model grid has also been developed
by \citet[]{Molinari08}, which focused on massive YSOs.  However, the
components in their model are relatively simple.  A massive YSO
involves complicated structures such as the protostar, accretion disk,
envelope and outflow, each of which may need multiple parameters to
define their properties that may affect the resulting SED.  Therefore,
to fit an observed SED usually requires setting a large number of
independent parameters. This is the method that the above mentioned
model grids have adopted. While the wide choice of free
parameters can generate good fits to the observations, such a method
usually also generates results that are physically less realistic or
not self-consistent (see, e.g., \citealt[]{Debuizer17}).  Large
numbers of free parameters will also lead to higher susceptibility to
degeneracies.

In this paper, the fourth of our series, we aim to use a different
approach to build the model grid. Instead of large numbers of free
parameters, we make the components more physically connected to reduce
the number of independent parameters.  Our model grid is based on a
particular model of massive star formation, the Turbulent Core model
(\citealt[]{MT02,MT03}).  In the Turbulent Core model, massive stars are formed
from pre-assembled massive pre-stellar cores, supported by internal
pressure that is provided by a combination of turbulence and magnetic
fields.  The pressure at the core surface is assumed to be
approximately the same as that of the surrounding larger-scale
star-cluster-forming clump, with a typical mean mass surface density
of $\scl\approx 1~\gcm$.  We construct the model grid from two initial
conditions, the initial core mass, $\mc$, and environmental mass
surface density, $\scl$.  With various analytical or semi-analytical
solutions, we calculate the properties of different components
including the protostar, disk, envelope, outflow and their evolutions
self-consistently from the initial conditions.  The main free
parameters in this model grid are the initial conditions, i.e., $\mc$
and $\scl$, and the protostellar mass $\ms$ indicating the evolutionary stage, as well as the inclination and
foreground extinction from the larger clump.  In such a method, the
model grid will exclude certain combinations of the components which
are not supported by the core accretion theory.  By fitting the
observed SEDs, this model will allow us to see whether the observed
variety of massive protostars can be explained by a scenario of core
accretion in different evolutionary stages and initial/environmental
conditions.

In the previous papers in our series (\citealt[]{ZT11}, hereafter
Paper I; \citealt[]{ZTM13}, hereafter Paper II; \citealt[]{ZTH14},
hereafter Paper III), we studied a fiducial case of a massive
protostar growing inside a core with an initial mass of $60~M_\odot$
and in a $1~\gcm$ environment, and a few variants of it.  Now in this
paper, we present the full model grid covering a large parameter
space, and investigate how the initial conditions and evolution affect
the SEDs of massive protostars (\S2).  We develop an SED fitting tool
to fit observed SEDs with this model grid (\S3).  In \S4, we
demonstrate the fitting process and results using the SED of the
massive protostar G35.20-0.74 as an example. We discuss our results
and present conclusions in \S5.

\section{Model Grid}

\subsection{Physical Model}
\label{sec:model}

We first briefly describe the physical assumptions used in our
models, which have been introduced in the
previous papers in this series. For detailed derivation and discussion
of these points, please refer to Papers I, II and III.  Following
\citet[]{MT03}, a star-forming core is defined as a region of a
molecular cloud that forms a single star or a close binary via
gravitational collapse. We can define such cores to contain a single,
central rotationally-supported disk. The initial core is assumed to be
quasi-spherical, self-gravitating, in near virial equilibrium, and in
pressure equilibrium with the surrounding star-cluster forming clump.
The size of such a core is determined by the mean mass surface density
of the surrounding clump $\scl$ (which sets the pressure on the
boundary of the core) by
\begin{equation}
R_c=5.7\times 10^{-2} (M_c/60\:M_\odot)^{1/2}(\scl/\gcm)^{-1/2}\:\mathrm{pc}
\end{equation}
(\citealt[]{MT03}). In the following text, $\scl$ is also referred to
as the mass surface density of the star-forming environment.  The
density distribution in the initial core is described by a power law
in spherical radius, $\rho\propto r^{-k_\rho}$. Observations suggest
$k_\rho$ has a mean value of 1.3 to 1.6 (\citealt[]{BT12};
\citealt[]{BTK14}). 
Therefore we adopt a fiducial value of $k_\rho=1.5$ for the whole model grid,
which is also consistent with our previous studies and the Turbulent Core Model by
\citet[]{MT03}.

The collapse of the core is described by an inside-out solution
(\citealt[]{Shu77}; \citealt[]{MP96,MP97}), together with the effect
of rotation (\citealt[]{Ulrich76}). Disks around massive protostars
are also expected to be massive due to the high accretion rates
of about $10^{-4}$ to $10^{-3}~\msunyr$ (e.g., \citealt[]{Beltran16b}).  
We assume the mass ratio between
the disk and the protostar is a constant $f_d=m_d/m_*=1/3$,
considering the rise in effective viscosity due to disk self-gravity
at about this value of $f_d$ (\citealt[]{Kratter08}).  Disk size is
calculated from the rotating collapse of the core to be $r_d(M_{*d}) =
0.684\:\beta_c\left(M_{*d}/m_{*d}\right)\left(M_{*d}/M_c\right)^{2/3}R_c$,
where $M_{*d}$ is the mass of the star-disk system in the limit of no
feedback as calculated from the collapse solution, and $m_{*d}$ is the
actual mass in the star-disk system (see Paper III).  The
rotational-to-gravitational energy ratio of the initial core $\beta_c$
is assumed to be 0.02, which is a typical value from observations of
low and high-mass prestellar cores (e.g. \citealt[]{Goodman93};
\citealt[]{Li12}; \citealt[]{Palau13}). The disk structure is
described with an ``$\alpha$-disk'' solution (\citealt[]{SS73}), with
an improved treatment to include the effects of the outflow and the
accretion infall to the disk (Paper II).  Half of the accretion energy
is released when the accretion flow reaches the stellar surface (i.e.,
the boundary layer luminosity,
$L_\mathrm{acc}=Gm_*\dot{m}_*/(2r_*)$). We assume this part of the
luminosity is radiated together with the intrinsic stellar luminosity
isotropically as a single black-body, i.e., the total luminosity from
the protostar is $L_{*,\mathrm{acc}}=L_*+L_\mathrm{acc}$ and the
surface temperature of the protostar is
$T_{*,\mathrm{acc}}=[L_{*,\mathrm{acc}}/(4\pi r_*^2\sigma)]^{1/4}$.
The other half of the accretion energy is partly radiated from the
disk during accretion and partly converted to the kinetic energy of
the disk wind. The total amount and the detailed distribution of the
accretion energy radiated from the disk are simultaneously derived
from the disk solution.

The density distribution of the disk wind is described by a
semi-analytic solution, which is approximately a \citet[]{BP82} wind
(see Appendix B of Paper II), and the mass loading rate of the wind
relative to the stellar accretion rate is assumed to be
$f_w=\dot{m}_w/\dot{m}_*=0.1$, which is a typical value for disk winds
(\citealt[]{KP00}). Such a disk wind carves out polar cavities in the
core, which gradually open up as the protostar evolves. The opening
angle of the outflow cavity is estimated following the method of
\citet[]{MM00} by comparing the wind momentum and that needed to
accelerate the core material to its escape velocity (Paper III).  The
accretion rate to the protostar is regulated by such outflow feedback.
Note that we allow existence of dust in some regions of the outflow
cavity if the disk winds in these regions originate from the disk
outside of the dust sublimation front.

The evolution of the protostar is solved using the model by
\citet[]{Hosokawa09} and \citet[]{Hosokawa10}.  The model solves
the detailed internal structure of the protostar, such as the
deuterium burning region, convective zone, and radiative zone, from
the accretion history calculated above (see Paper III for more
details). A photospheric boundary condition, which is usually
associated with the situation of disk accretion, is used in the
protostellar evolution calculation. Several outputs of this
calculation that are important for setting up our grid of physical
models for radiative transfer computation include the evolution of
protostellar radius, luminosity and surface temperature with the
protostellar mass.

\begin{figure}
\begin{center}
\includegraphics[width=\columnwidth]{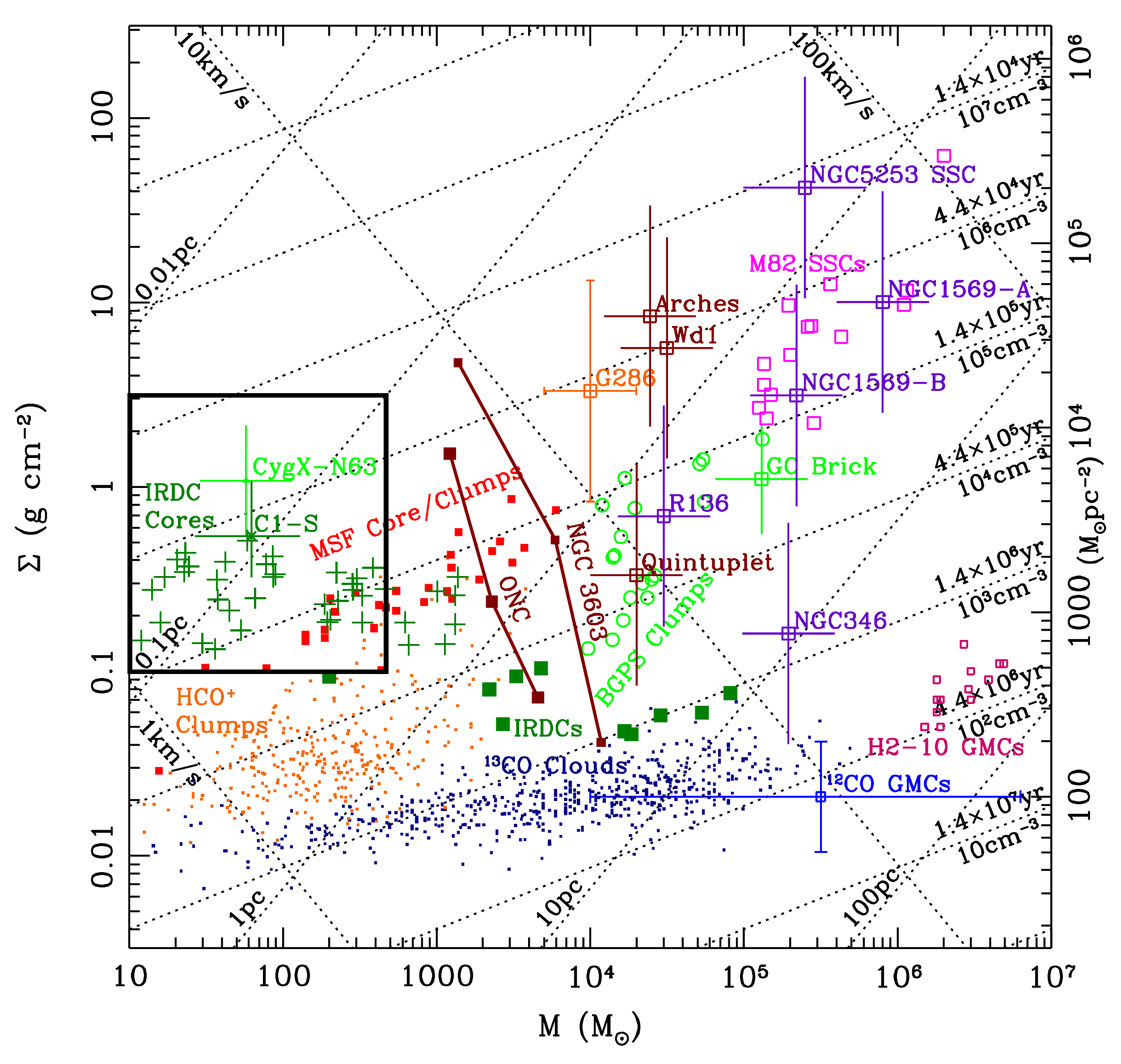}
\caption{The range of initial conditions of the cores in the model
  grid (thick black rectangle) compared with the observed environments
  of massive star formation. The latter is taken from Figure 1 of
  \citet[]{Tan14}, showing masses and mass surface densities of
    example GMCs, massive gas clumps, Infrared Dark Clouds and their
    internal clumps/cores, and young star clusters (see also similar
    figures with more recent data added in, e.g., \citet[]{Elia17}).}
\label{fig:Mc-Sigma}
\end{center}
\end{figure}

\begin{figure}
\begin{center}
\includegraphics[width=0.5\columnwidth]{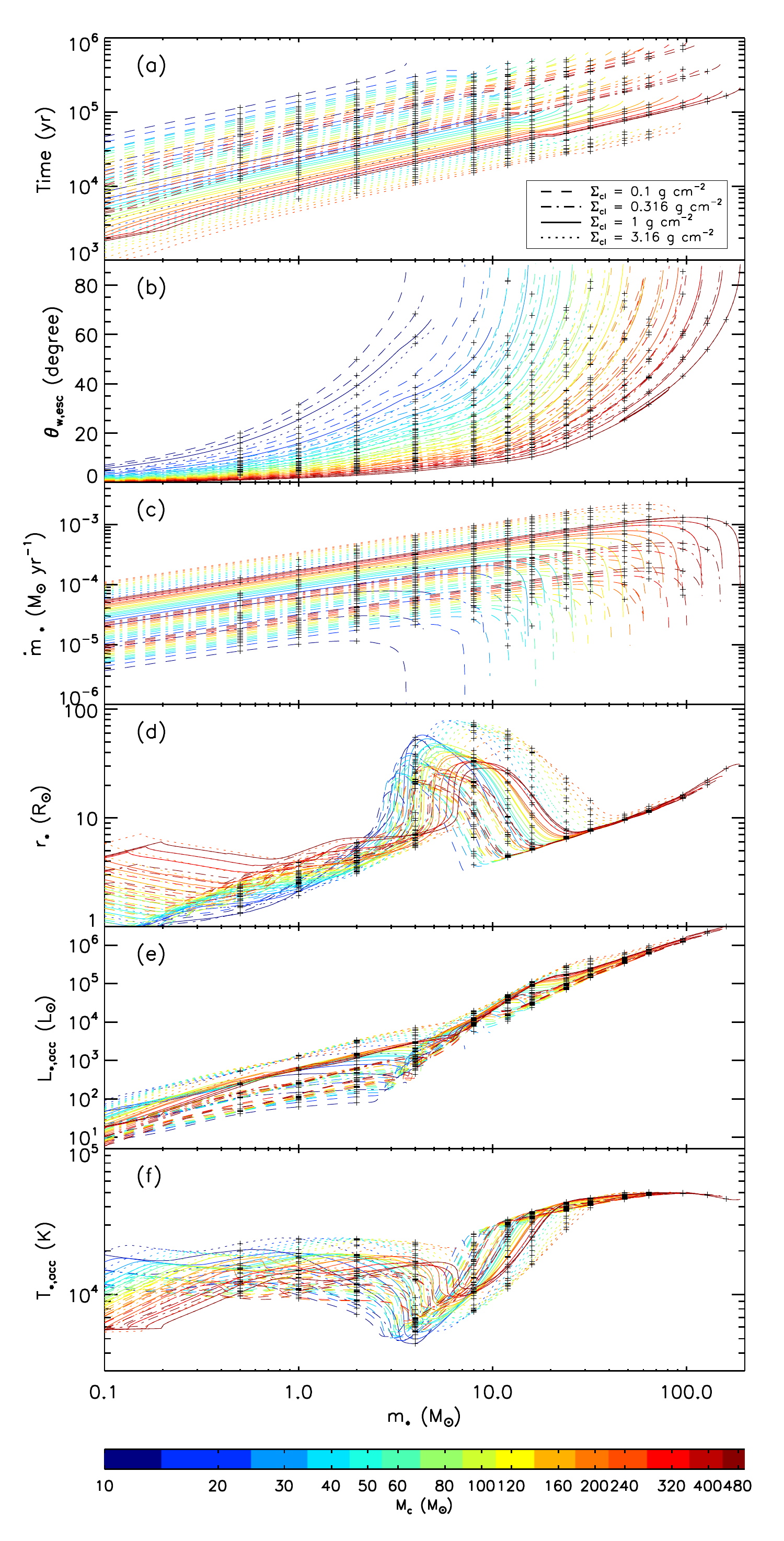}
\caption{Evolutionary tracks in the model grid versus protostellar mass,
$\ms$. {\bf (a) $-$ (f):} Time since formation of the protostar, i.e.,
protostellar age (a); half opening angle of outflow cavity $\thetaw$ (b);
protostellar accretion rate $\dot{m}_*$ (c); protostellar radius $r_*$ (d);
protostellar luminosity $L_{*,\mathrm{acc}}$ (e); and protostellar surface
temperature $T_{*,\mathrm{acc}}$ (f), with growth of the protostellar mass.
Evolutionary tracks for different
initial core masses, $\mc$, are shown in different colors and
different clump environment mass surface densities, $\scl$, are shown
by different line styles. The cross symbols mark the sampling of
models, i.e., for which radiative transfer calculations are performed,
that form the model grid.}
\label{fig:history}
\end{center}
\end{figure}

%

\begin{figure}
\begin{center}
\includegraphics[width=0.5\columnwidth]{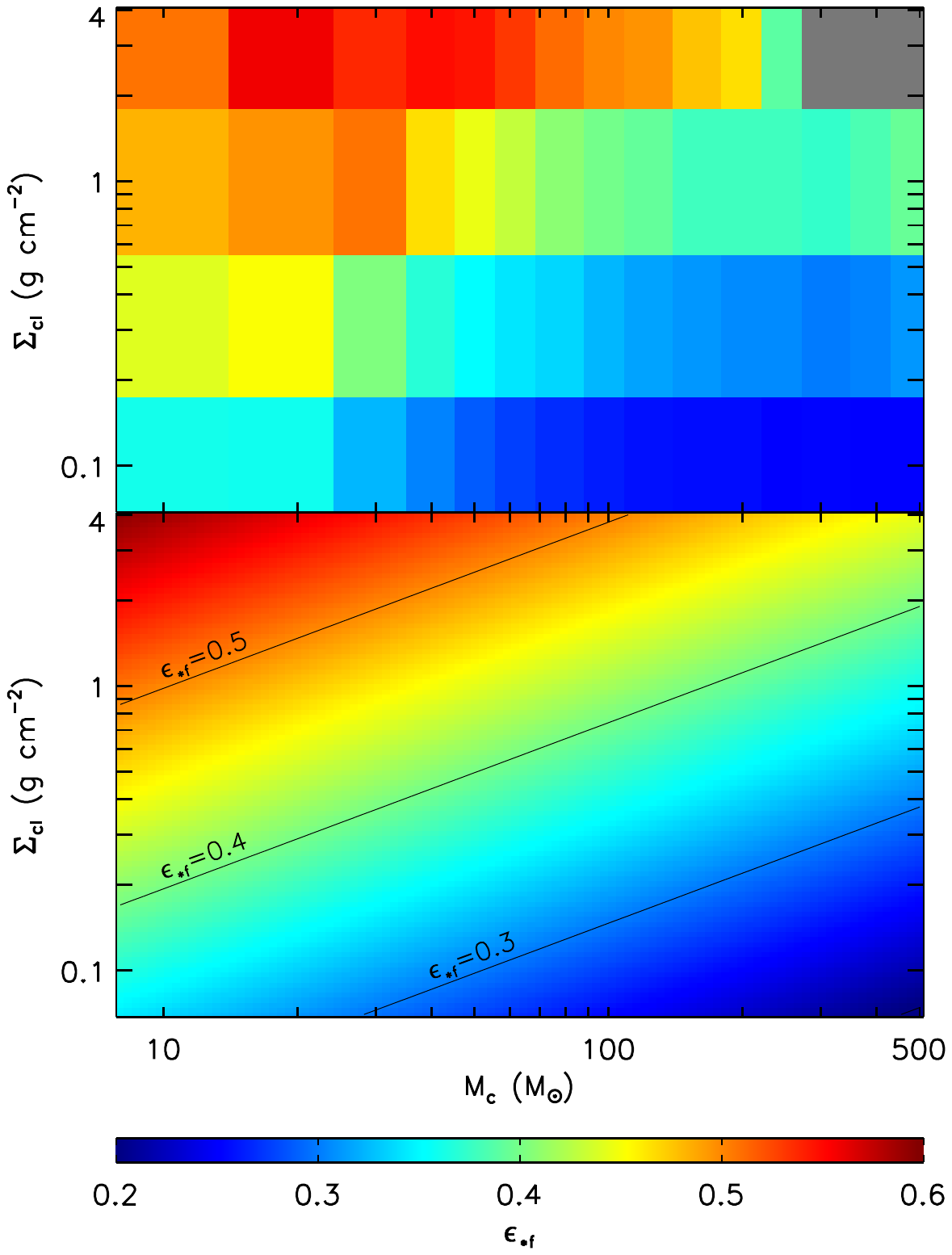}
\caption{Dependence of the star formation efficiency, $\epsilon_{*f}\equiv
m_{*f}/M_c$ (shown in color scale), on the initial core mass, $M_c$,
and the mass surface density of the ambient clump, $\scl$.  The
efficiencies are calculated for each evolutionary track in the model
grid (upper panel) and then expanded via a two dimension linear
regression to the whole initial condition parameter space, i.e.,
$\log\mc-\log\scl-\epsilon_{*f}$ {\bf (lower panel)}. The $\epsilon_{*f}$ data are missing
for models with most massive cores in highest $\scl$ environments
(grey area in upper-right of upper panel), due to the difficulty of
the protostellar evolution calculation at very high accretion rates of
reaching the final stellar mass.  The contours in the lower panel are
$\epsilon_{*f}=0.3$, 0.4, and 0.5.}
\label{fig:eff}
\end{center}
\end{figure}

\begin{figure}
\begin{center}
\includegraphics[width=\columnwidth]{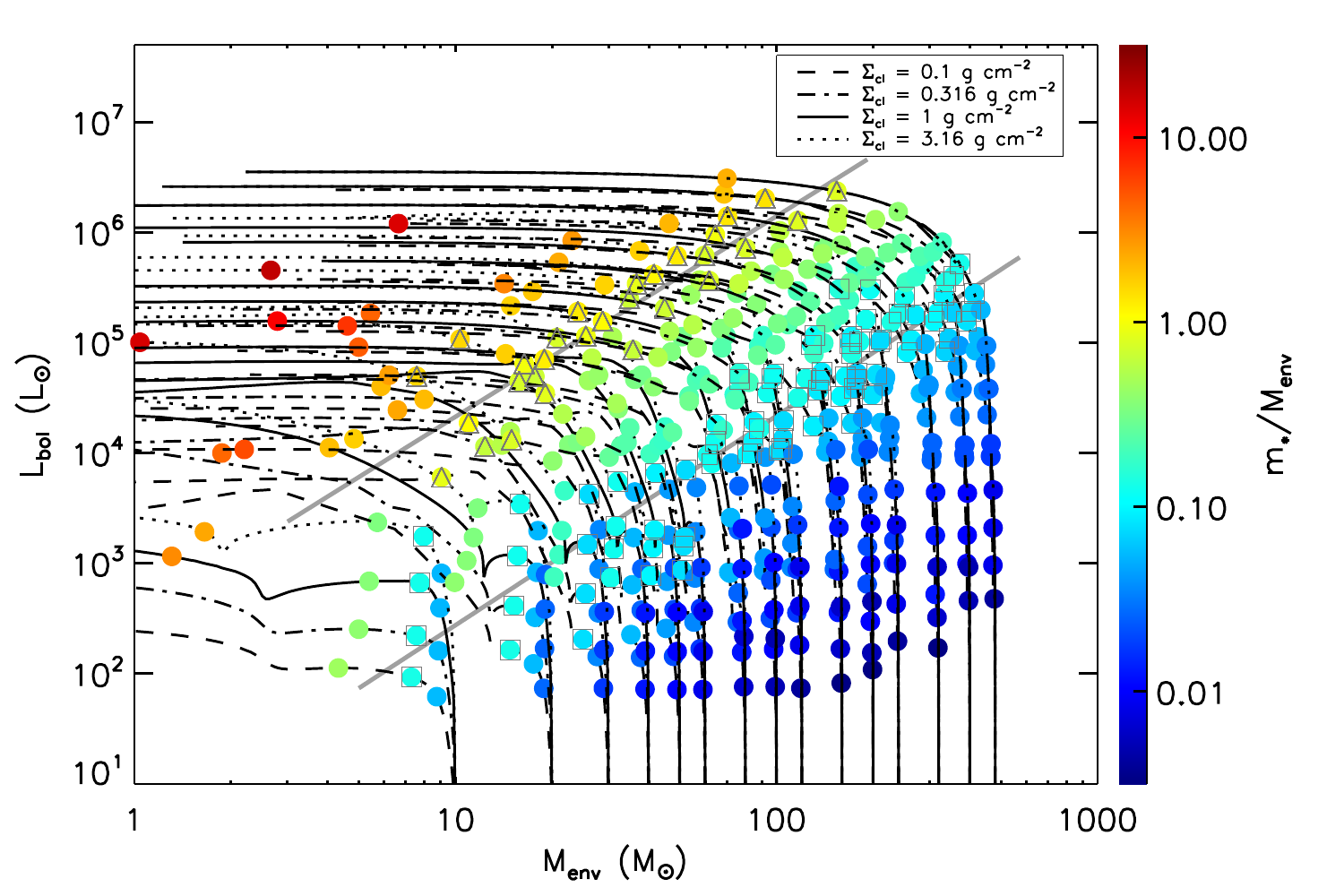}
\caption{Evolution of the bolometric luminosity with envelope mass along the
evolutionary tracks of the model grid. The circles mark the
evolutionary stages for which we perform radiative transfer
simulations to build the model grid. The color of the circles indicate
the evolutionary stages expressed as $\ms/\menv$. The open triangles
and squares mark the evolutionary stages with
$-0.2<\log(\ms/\menv)<0.2$ and $-1.2<\log(\ms/\menv)<-0.8$,
respectively. The grey straight lines are log-log linear fits to these
two groups of models.}
\label{fig:masslum}
\end{center}
\end{figure}

\subsection{Parameter Space}
\label{sec:parameter}

In such a framework, the evolution of the core, protostar, disk and
outflow cavity are self-consistently calculated from two main initial
conditions of the core: its initial mass ($\mc$) and the mean mass
surface density of the clump that the core is embedded in ($\scl$).
We refer to the evolutionary history of protostar from a given set of
initial conditions as an {\it evolutionary track}, and a particular
moment on such a track as an {\it evolutionary stage}, which is
specified by a third parameter, the protostellar mass $\ms$.  We refer
to the entire set of tracks as the {\it model grid}.  Therefore, this
model grid is of three dimensions ($\mc-\scl-\ms$). In the current
model grid, $\mc$ is sampled at 10, 20, 30, 40, 50, 60, 80, 100, 120,
160, 200, 240, 320, 400, 480~$M_\odot$, and $\scl$ is sampled at 0.1,
0.32, 1, 3.2~$\gcm$, forming 60 evolutionary tracks. $\ms$ is sampled
at 0.5, 1, 2, 4, 8, 12, 16, 24, 32, 48, 64, 96, 128, 160 $M_\odot$.
Note that for each track, not all of these $\ms$ are sampled.  In
particular, the maximum protostellar mass is limited by the final
stellar mass achieved in a given evolutionary track (see Figure
\ref{fig:eff}). As a result, there are totally 432 different
physical models defined by different sets of $(\mc,~\scl,~\ms)$.

We note that other initial conditions may affect the models, such as
the initial rotational-to-gravitational energy ratio of the core $\beta_c$ 
and magnetic field strength in the core, both of which are
expected to influence the size of the accretion disk. However, the
spectral energy distributions are not significantly affected by
variations of the disk size about its fiducial values, given that the
disk is always large compared to the size of the star and small
compared to the extent of the core and outflow cavities (see Paper III).

Figure \ref{fig:Mc-Sigma} shows the range of the initial conditions in
our model grid.  The mass surface density of the star-forming
environment, $\scl$, covers a range from $0.1$ to $3.2~\gcm$. This
range is similar to the values found in most Galactic massive star
forming regions, including: infrared dark clouds (IRDCs, dark green
squares) and their internal clumps/cores (dark green crosses) that are
thought to represent the initial stage of massive star formation;
massive star-forming clumps/cores (red squares and light green
circles), including those at Galactic Center (e.g., the ``Brick'');
some massive star clusters (e.g., the Orion Nebula Cluster [ONC]) and
even more massive ``super star clusters'' (e.g., Westerlund 1, Arches,
and Quintuplet).  The initial core mass, $\mc$, in the model grid
covers a range from $10$ to about $500~M_\odot$, which is similar to
those of individual pre-stellar and protostellar cores inside IRDCs
and massive clumps.  Therefore, our model grid covers a wide range of
initial conditions that are suitable to form individual stars from
intermediate to high-mass.  We note that as the first release of
  the model grid, the current version does not have very fine sampling
  over the initial conditions, especially the surface density of the
  star-forming environment $\scl$. While the sampling of $\scl$ in the
  current model grid, which covers most of the relevant range of local
  massive star formation, is sufficient to understand the differences
  between low and high $\scl$ models, as we will show in the examples
  of SED model fitting in \S\ref{sec:example}, there can be
  degeneracies that span the full range of $\scl$.  Therefore we note
  that the constraints placed on $\scl$ at this point are still be
  relatively limited.

Figure \ref{fig:history} shows all the evolutionary tracks in the
current model grid.  Panel (a) shows how the growth of the
protostellar mass corresponds to time, i.e., protostellar age.  The
evolutionary stages we sample in the grid (marked by the crosses)
cover a range of age from about $7\times 10^3$ yr to about
$7\times10^5$ yr.  Panels (b) to (f) show the evolution of the outflow
cavity opening angle, accretion rate, protostellar radius, luminosity
and surface temperature, with the growth of protostellar mass.  All
the models show similar trends in these figures.  As the protostar
grows, the outflow cavity gradually opens up (Panel b) due to the
interaction between the outflow and the core, i.e., outflow
feedback. The accretion rate increases with protostellar mass for most
of the time, except at the final stages when outflow feedback causes
the accretion rate to decrease (Panel c).  The moments when the
accretion rates start to decline happen around $0.5 - 0.9$ of the
total formation times, and up to such moments, the protostars have
grown to $0.5 - 0.8$ of their final masses.  The protostellar
evolutions can be divided into several stages, which are clearly seen
in the change of protostellar radius (Panel d).  Note that the
  radius is calculated from the protostellar evolution model
  (\S\ref{sec:model}), which then helps determine the photospheric
  properties of the protostar.  At the lowest protostellar masses
that the model grid covers ($\sim0.5\:M_\odot$), the protostellar
radius steadily grows with mass due to deuterium burning. From about
several $\times M_\odot$, the radius increases drastically with
protostellar mass, caused by the redistribution of entropy in the
protostar. The radius reaches its peak at $m_*=4 - 10\:M_\odot$, after
which the protostar enters the Kelvin-Helmholz (KH) contraction
stage. The main-sequence stage starts from $\gtrsim 10\:M_\odot$ in
the low $\scl$ cases and from $\gtrsim 30\:M_\odot$ in the high $\scl$
cases.  The luminosity $L_{*,\mathrm{acc}}$, which combines the
intrinsic protostellar luminosity and the boundary-layer accretion
luminosity, almost monolithically increases with protostellar mass
(Panel e). It is worth noting that the accretion luminosity is
dominant before the fast swelling phase of the protostar, after which
the protostellar luminosity from nuclear (hydrogen and/or deuterium)
burning becomes dominant.  The surface temperature of the protostar is
significantly affected by protostellar radius, especially around the
fast-swelling phase and the KH contraction phase (Panel f).

While the general trends of the evolutionary tracks in the model grid
are similar, they are affected by the initial conditions, especially
the mass surface density of the environment $\scl$.  If the core of a
given mass is embedded in an environment with a higher surface
density, then it is more pressurized and becomes more compact and
denser, and thus collapses more quickly, leading to a shorter star
formation timescale, a higher accretion rate, and a higher luminosity.
Such a core is also more resistant to outflow feedback, so the outflow
cavity opens up more slowly with $\ms$ compared with a core in a low $\scl$
environment.  In such a case, with the higher accretion rate, the
protostar also enters the fast swelling phase and main-sequence stage
at a higher mass, and reaches a larger radius during the swelling
phase, which affects the stellar surface temperature.

Figure \ref{fig:eff} shows the star formation efficiencies from the
initial cores to the final stars, $\epsilon_{*f}\equiv m_{*f}/\mc$, in
the model evolutionary tracks.  Such efficiencies are calculated for each
evolutionary track and then expanded to the whole initial condition
parameter space via a two dimension linear regression (in log space).
The fitted relation between $\epsilon_{*f}$ and the initial conditions
is
\begin{equation}
\epsilon_{*f}=0.44-0.083\log\left(\frac{\mc}{60~M_\odot}\right)+0.14\log\left(\frac{\scl}{1~\gcm}\right),
\end{equation}
which agrees with the data points (the upper panel of Figure
\ref{fig:eff}) within 17\%.  In the model grid, the star formation
efficiency ranges from about 0.2 to about 0.6.  Such values are
consistent with the scenario that the stellar initial mass function
(IMF) is inherited from the core mass function (CMF) with a relatively
constant core-to-star conversion efficiency (e.g.,
\citealt[]{Alves07}; \citealt[]{Andre10}; \citealt[]{Offner14}), but
now somewhat dependent on the distribution of $\scl$ for the global
population of pre-stellar cores.

For a core of a given mass, the star formation efficiency is higher in
an environment with a higher mass surface density. This is because
cores are denser in a higher surface density environment and so it is
harder for the outflow to disperse the envelope and stop the accretion.  On the
other hand, in the same environment, the star formation efficiency
decreases as the core mass increases. This is expected since the
feedback becomes stronger as the protostar grows.  In our model grid,
we only include mechanical feedback from the outflow momentum, while
ignoring other feedback mechanisms such as radiation pressure,
photoevaporation, and stellar winds. For several fiducial models in our
model grid, \citet[]{Tanaka17} have included these effects and
found that while the mechanical feedback from the outflow is always
dominant, the other effects especially radiation pressure can
significantly affect the star formation efficiency for the most
massive cores in our model grid (forming $> 100\:M_\odot$ stars) in
low mass surface density environments ($\scl\lesssim
0.3\:\gcm$). Following this trend, the relatively low efficiencies in
the lower-right corner of Figure \ref{fig:eff} will become even lower.
However, \citet[]{Tanaka17} did not find any sudden decrease of the
efficiency with core mass at the high-mass end, even with the
additional feedback included, suggesting that such feedback does not
lead to a truncation of the high-end of the stellar IMF.

Figure \ref{fig:masslum} shows the evolutionary tracks of the model
grid in the $\lbol-\menv$ plane.  $\menv$ is the current envelope
  mass, which is different from the initial mass of the core $M_c$. As
  the protostar evolves, $\menv$ gradually decreases due to the
  accretion to the protostar and widening of the outflow cavity.  The
$\lbol-\menv$ diagrams have been used to identify the evolutionary
stages of massive protostars (e.g., \citealt[]{Molinari08};
\citealt[]{Elia10}; \citealt[]{Konig17}).  As expected, the model
evolutionary tracks start in the lower-right of the figure and
gradually move to the upper-left.  Although the models cover a wide
range of $\lbol$ and $\menv$, models with similar evolutionary stages
(defined by $\ms/\menv$) form strips on this diagram.  We fit two
groups of models at different evolutionary stages. One group is around
$\ms/\menv=1$, which is usually used to mark the end of the main
accretion phase and the start of the envelope clear-up phase, and
corresponds to the Class 0 to Class I transition in low-mass star
formation.  Another group is around $\ms/\menv=0.1$, i.e., typical
sources in the main accretion phase.  The fitting results to these two
groups are $\log(L_\mathrm{bol}/L_\odot)=2.5+1.8\log(\menv/M_\odot)$
and $\log(L_\mathrm{bol}/L_\odot)=0.54+1.9\log(\menv/M_\odot)$, which
have similar slopes but one is 2 order of magnitude below the other in
$\lbol$.  Compared with the results of similar fits to the observed
data by \citet[]{Molinari08} and \citet[]{Konig17}, the slopes
predicted by our models are steeper than the observations, which are
in the range $0.5-1.3$.  It is worth noting that only the core mass is
included in our model, while in real observations, additional clump
material will usually be included due to the low resolutions of
single-dish FIR observations.  This effect will tend to cause the
observed slopes to be shallower, if more luminous, typically more
distant, sources tend to have more contamination from surrounding
clump material.  However, \citet[]{Baldeschi17} found that
  increasing distance, although causing source positions to shift in
  the $L-M$ diagram, does not change the slope, on average.  In
addition, the luminosity used in this diagram is the true total
luminosity from the source, which may differ from the luminosity
directly integrated from the SEDs by an order of magnitude or more
depending on the viewing inclination, since more radiation is emitted
in the polar direction due to low density outflow cavities (see
\S\ref{sec:flashlight}).

\subsection{Radiative Transfer Simulations and Resultant SED Grid}
\label{sec:sedgrid}

The Monte-Carlo continuum radiative transfer simulation is performed
for the protostellar cores in the model grid using the HOCHUNK3d code
by \citet[]{Whitney03,Whitney13}.  Different dust opacity models
  are assigned to different regions, including the envelope, the low
  density regions of the disk ($n_\mathrm{H} < 2
  \times10^{10}~\mathrm{cm}^{-3}$), the high density regions of the
  disk, the part of outflow launched from the disk outside of the dust
  sublimation radius, and the foreground ISM (for calculating the
  foreground extinction; see below).  Details about these opacity
  models were described in Paper I and II.  The dust models and setups
  are same as those used by \citet[]{Robitaille06,Robitaille07}.  The
code was updated to also include gas opacities (which is important in
high temperature regions around massive protostars), adiabatic
cooling/heating and advection (Papers I \& II).  For each model, the
temperature profile is calculated and SEDs at 20 viewing inclinations
are produced. The inclination is sampled at
$\muinc\equiv\cos\inc=0.975,~0.925,~...~,~0.025$, i.e., equally
distributed between 1 (face-on) and 0 (edge-on).  Therefore there are
in total 8640 SEDs in the current model grid determined by 4
independent parameters $\mc$, $\scl$, $\ms$, and $\inc$.  Note that
these SEDs include all the emission from the source, i.e., with an
aperture which is large enough to cover the whole core.

In order to compare with the observation, the model SEDs need first to
be scaled by the distance, and then adjusted by additional foreground
extinction described by the parameter $\av$,
\begin{equation}
\fnumodav(\lambda)=\fnumod(\lambda)\times 10^{-0.4\av\kappa(\lambda)/\kappa_V}, \label{eq:fnumodav}
\end{equation}
where $\kappa(\lambda)$ and $\kappa_V$, the dust opacities at the
wavelengths $\lambda$ and in the $V$-band, respectively, are from the extinction law
of the dust model, and $\fnumod$ are the distance-scaled model fluxes.
When fitting actual data (see \S\ref{sec:fitting}), the model SEDs are
further convolved with the transmission profiles of the instrument
filters to simulate the fluxes detected in observational bands of
various instruments.
Therefore, in principle, it is not necessary to perform color correction to the observed fluxes
before fitting with the model grid.

In such a model grid, we are explicitly linking the SEDs to the
initial conditions and evolutionary stages of massive star
formation. An SED to fit the observation is determined by six parameters: $\mc$, $\scl$,
$\ms$, $\inc$, $d$, and $\av$.  Such an approach assumes different
components are physically connected to each other, therefore reducing
the dimensionality of the parameter space and thus the number of model
that need to be computed.  Meanwhile, by comparing such models with
observations, especially through fitting a large sample of massive
protostars, one can understand to what extent the observed variety in
the infrared continuum emission of massive protostars can be explained
simply by different initial conditions and evolutionary stages, and
ultimately test the turbulent core accretion theory of massive star
formation.

\subsubsection{Example SEDs}
\label{sec:sed_fiducial}

\begin{figure}
\begin{center}
\includegraphics[width=0.5\columnwidth]{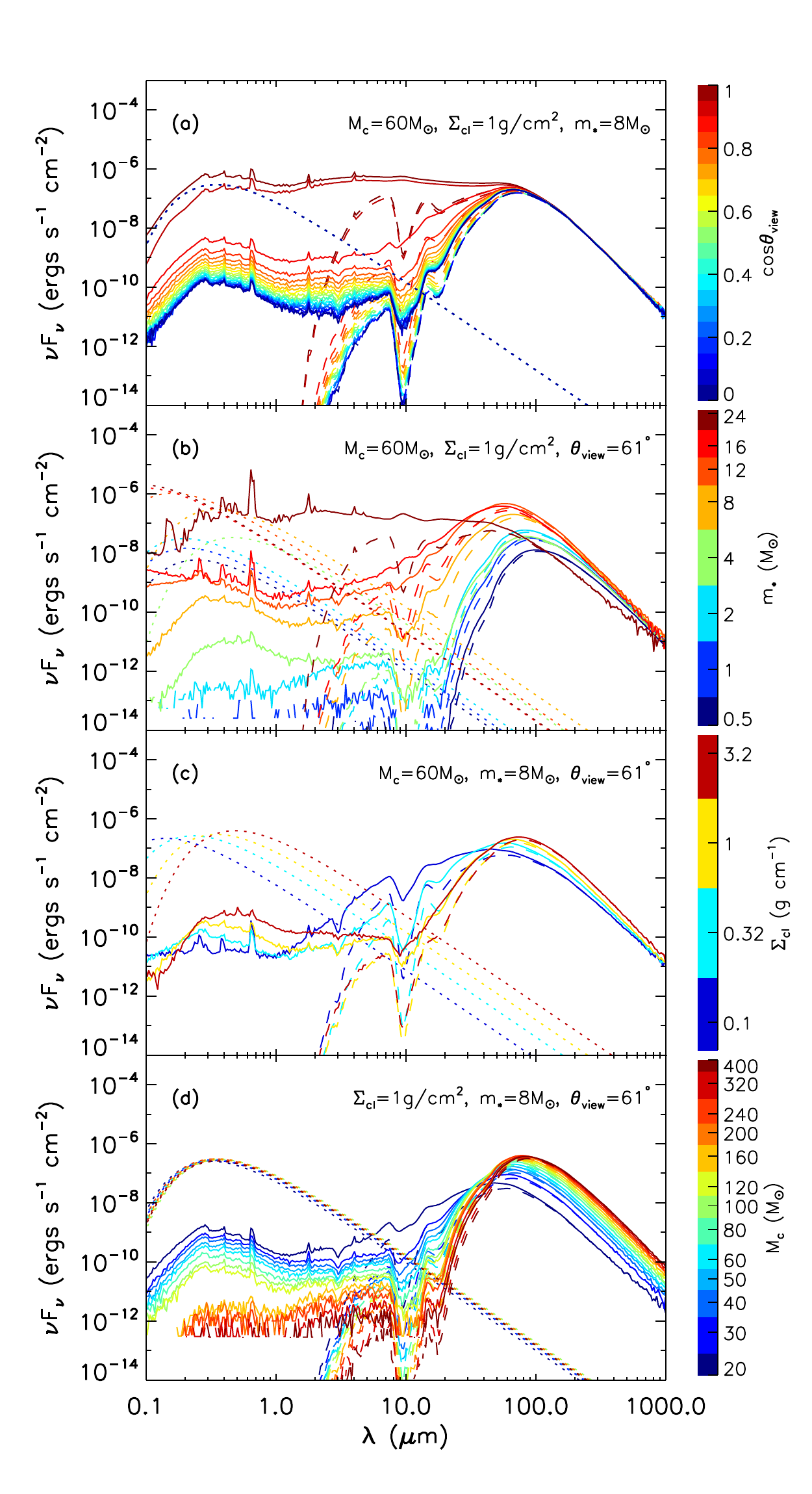}
\caption{Examples of the SEDs in the model grid. A distance of 1 kpc is
assumed.  In each panel, the solid lines and the dashed lines are the
model SEDs without any foreground extinction and with a foreground
extinction of $\av=100$, respectively.  The dotted lines are the input
protostellar SEDs (with boundary layer accretion luminosity included).
{\bf (a):} The SEDs of models with the same $\mc=60\:M_\odot$,
$\scl=1\:\gcm$ and $\ms=8\:M_\odot$, but at different viewing
inclinations, which are shown by the color scale.  {\bf (b):} The SEDs
of models with the same $\mc=60\:M_\odot$, $\scl=1~\gcm$ and $\inc=
61^\circ$, but different values of $\ms$.  {\bf (c):} The SEDs of
models with the same $\mc=60\:M_\odot$, $\ms=8\:M_\odot$ and
$\inc=61^\circ$, but different values of $\scl$.  {\bf (d):} The SEDs
of models with the same $\scl=1\:\gcm$, $\ms=8\:M_\odot$ and
$\inc=61^\circ$, but different values of $\mc$.}
\label{fig:sed_fiducial}
\end{center}
\end{figure}

\begin{figure*}
\begin{center}
\includegraphics[width=\textwidth]{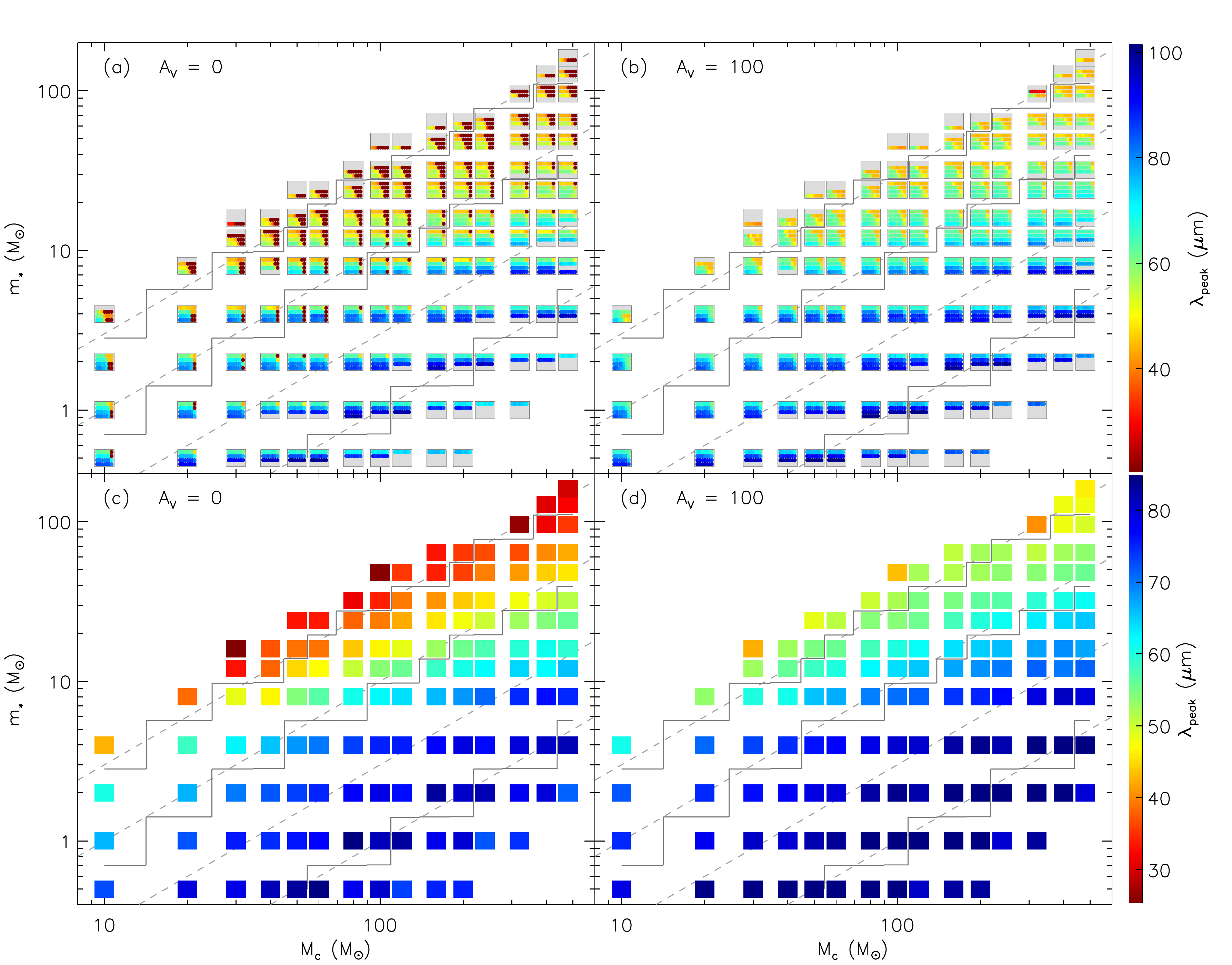}
\caption{Distribution of the SED peak wavelengths (shown in color scale)
in the parameter space of the model grid.  
{\bf (a):} Each small square is a group of models for each set of $\mc$
and $\ms$. Inside each square, the four rows from top to bottom are
$\scl=0.1$, 0.32, 1, and 3.2~$\gcm$, and each column from left to
right is the inclination angle $\inc$ from an edge-on view to a
face-on view.  {\bf (b):} Similar to Panel (a), but calculated from
SEDs with a foreground extinction of $\av=100$.  {\bf (c):} Similar to
Panel (a), but the color of each square shows the SED peak position
averaged over $\scl$ and $\inc$ at each $\mc$ and $\ms$.  {\bf (d):}
Similar to Panel (c), but calculated from SEDs with a foreground
extinction of $\av=100$.  The dashed lines are where $\ms/\mc=0.01$,
0.03, 0.1, and 0.3.  The solid lines are where $\ms/\menv=0.01$, 0.1,
and 1. Here $\menv$ is averaged over $\scl$ for each $\mc$ and $\ms$.}
\label{fig:sed_peak}
\end{center}
\end{figure*}

%

\begin{figure*}
\begin{center}
\includegraphics[width=\textwidth]{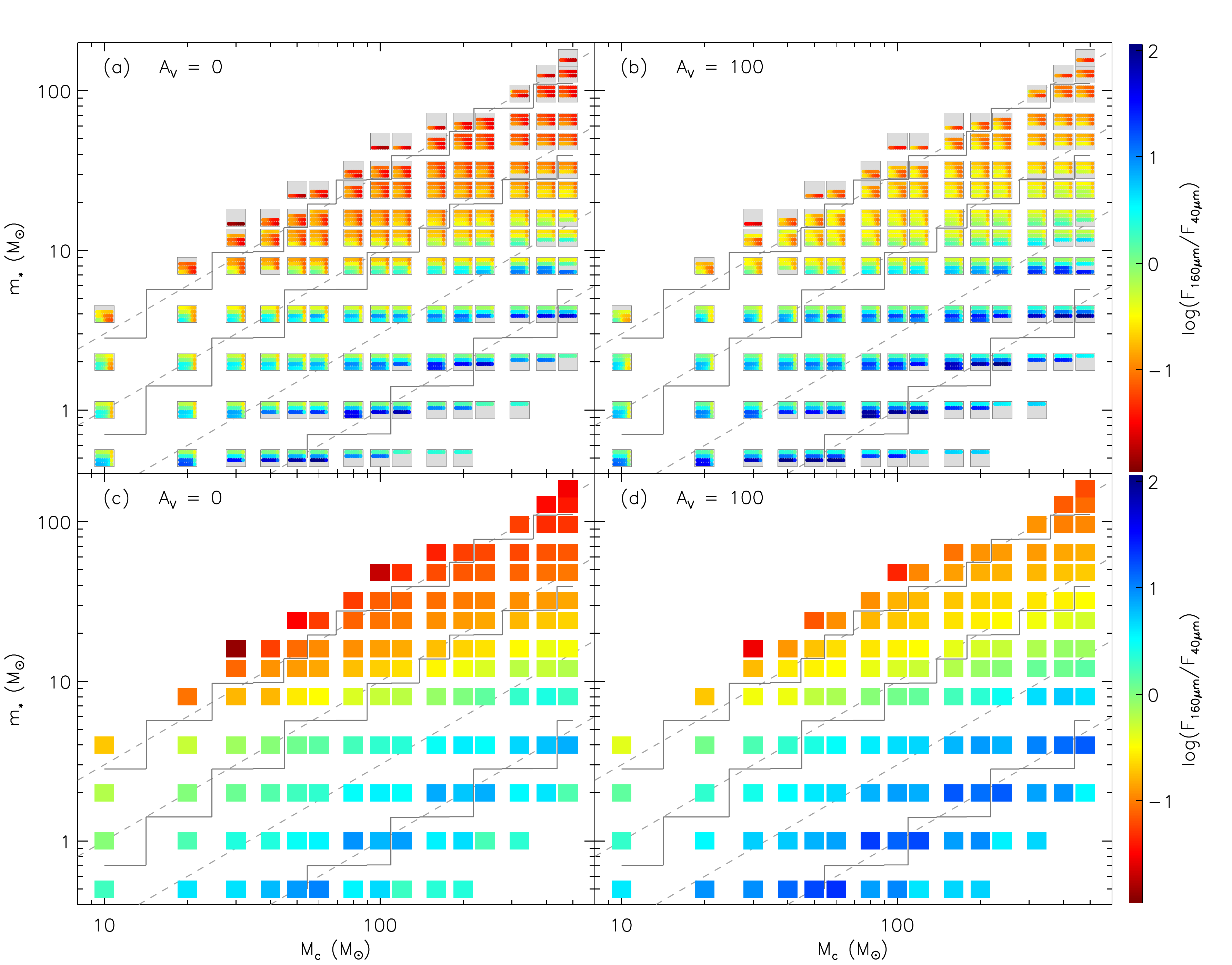}
\caption{Distributions of the [160~$\mu$m]$-$[40~$\mu$m] color (defined as
$\log[F_{\rm 160\mu m}/F_{\rm 40\mu m}]$, shown in color scale).  {\bf
  (a):} Each small square is a group of models for each set of $\mc$
and $\ms$. Inside each square, the four rows from top to bottom are
$\scl=0.1$, 0.32, 1, and 3.2~$\gcm$, and each column from left to
right is the inclination angle $\inc$ from an edge-on view to a
face-on view.  {\bf (b):} Similar to Panel (a), but calculated from
SEDs with a foreground extinction of $\av=100$.  {\bf (c):} Similar to
Panel (a), but each square shows the [160~$\mu$m]$-$[40~$\mu$m] color
averaged over $\scl$ and $\inc$ at each $\mc$ and $\ms$.  {\bf (d):}
Similar to Panel (c), but calculated from SEDs with a foreground
extinction of $\av=100$.  The dashed lines are where $\ms/\mc=0.01$,
0.03, 0.1, and 0.3.  The solid lines are where $\ms/\menv=0.01$, 0.1,
and 1. Here $\menv$ is averaged over $\scl$ for each $\mc$ and $\ms$.}
\label{fig:sed_midtofar}
\end{center}
\end{figure*}

\begin{figure*}
\begin{center}
\includegraphics[width=\textwidth]{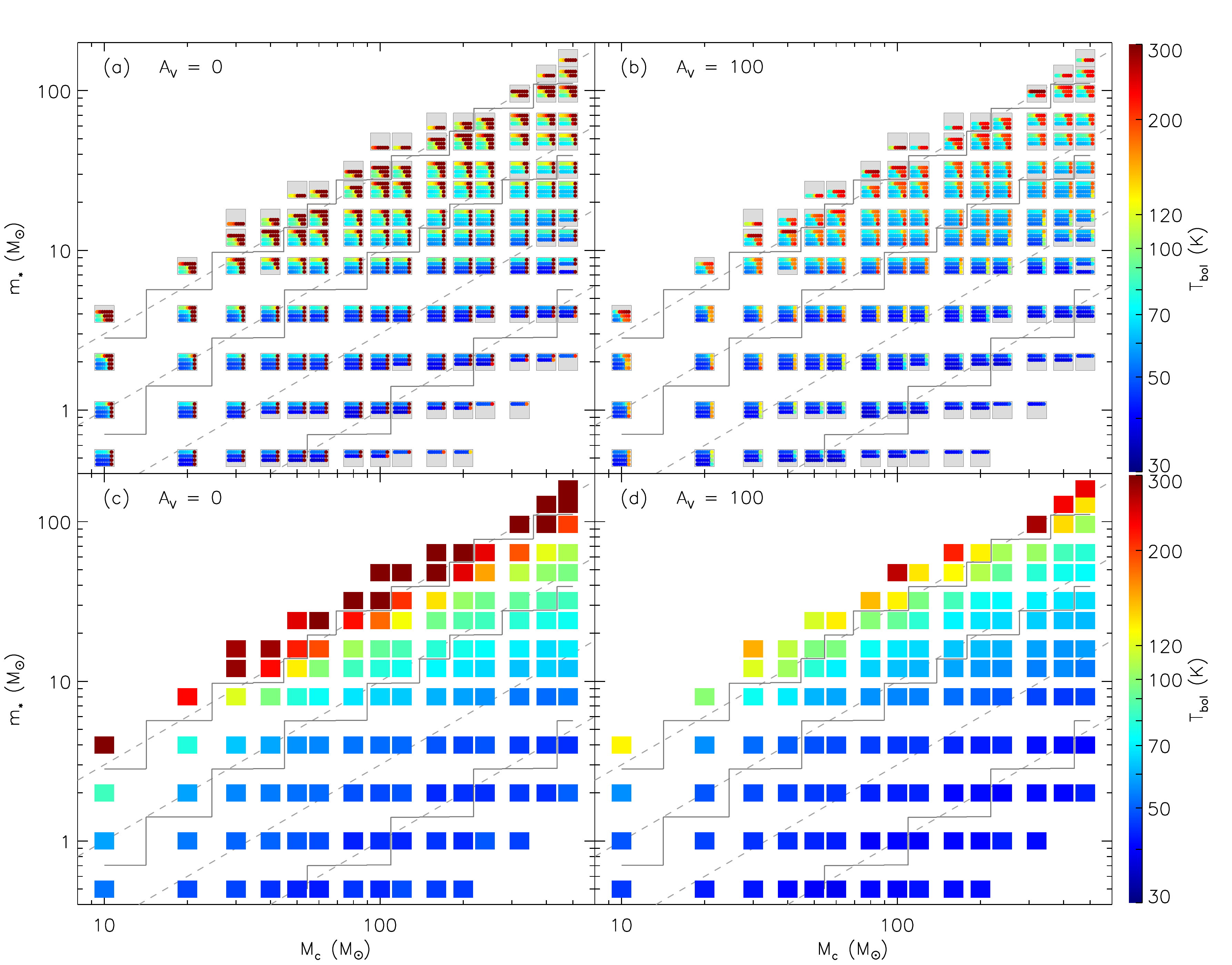}
\caption{Distributions of the bolometric temperature (shown in color scale).
{\bf (a):} Each small square is a group of models for each set of
$\mc$ and $\ms$. Inside each square, the four rows from top to bottom
are $\scl=0.1$, 0.32, 1, and 3.2~$\gcm$, and each column from left to
right is the inclination angle $\inc$ from an edge-on view to a
face-on view.  {\bf (b):} Similar to Panel (a), but calculated from
SEDs with a foreground extinction of $\av=100$.  {\bf (c):} Similar to
Panel (a), but the color of each square shows the bolometric
temperature at inclination angle of $60^\circ$ and averaged over $\scl$ at
each $\mc$ and $\ms$.  {\bf (d):} Similar to Panel (c), but calculated
from SEDs with a foreground extinction of $\av=100$.  The dashed lines
are where $\ms/\mc=0.01$, 0.03, 0.1, and 0.3.  The solid lines are
where $\ms/\menv=0.01$, 0.1, and 1. Here $\menv$ is averaged over
$\scl$ for each $\mc$ and $\ms$.}
\label{fig:tbol}
\end{center}
\end{figure*}


\begin{figure*}
\begin{center}
\includegraphics[width=0.9\textwidth]{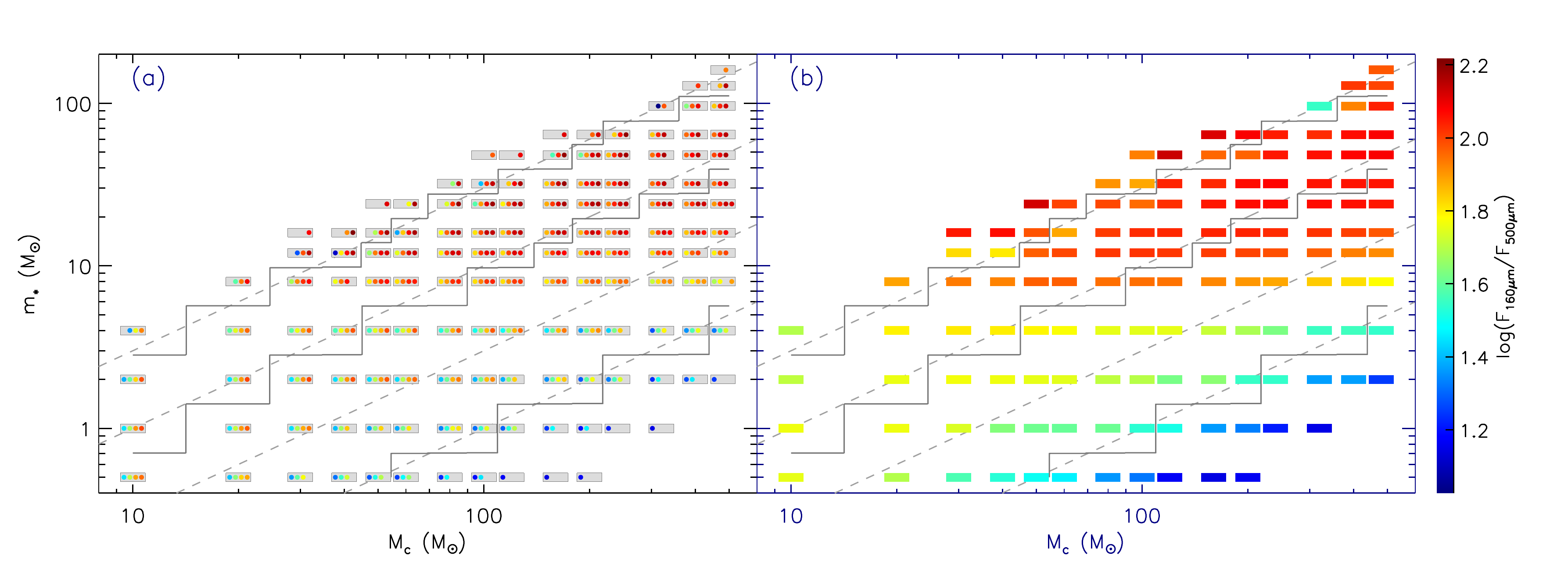}
\caption{{\bf (a):} Distribution of the [160~$\mu$m]$-$[500~$\mu$m] color
(defined as $\log[F_{\rm 160\mu m}/F_{\rm 500\mu m}]$, shown in color
scale). Each rectangle is a group of models for each set of $\mc$ and
$\ms$.  The four points inside each rectangle from left to right are
$\scl=0.1$, 0.32, 1 and 3.2~$\gcm$. At this wavelength range, the SEDs
are not affected by the inclination or the foreground extinction.
{\bf (b):} Similar to Panel (a), but each rectangle shows the
[160~$\mu$m]$-$[500~$\mu$m] color averaged over $\scl$ at each $\mc$
and $\ms$.  The dashed lines are where $\ms/\mc=0.01$, 0.03, 0.1, and
0.3.  The solid lines are where $\ms/\menv=0.01$, 0.1, and 1. Here
$\menv$ is averaged over $\scl$ for each $\mc$ and $\ms$.}
\label{fig:sed_far}
\end{center}
\end{figure*}

\begin{figure*}
\begin{center}
\includegraphics[width=0.9\textwidth]{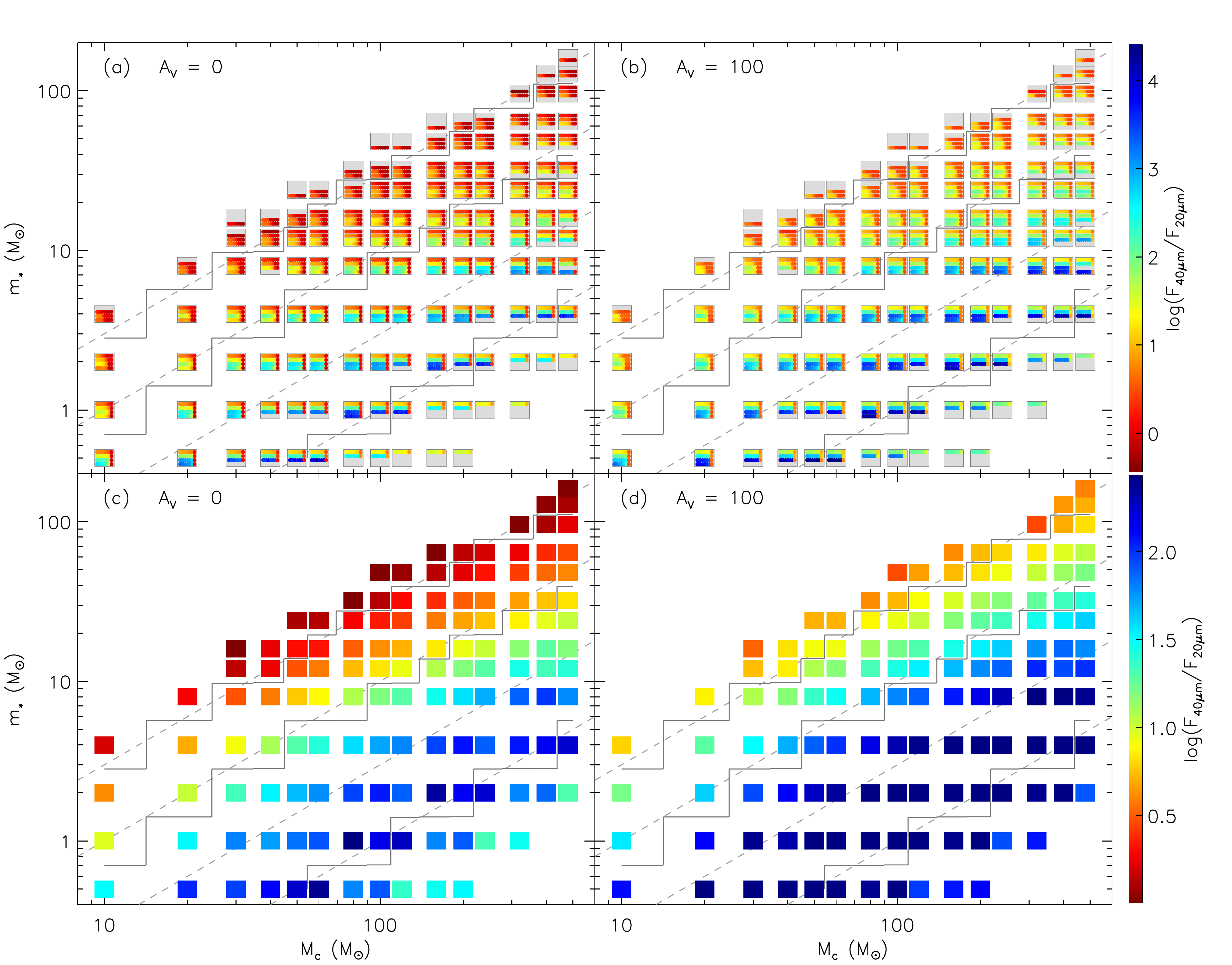}
\caption{Distributions of the [40~$\mu$m]$-$[20~$\mu$m] color (defined as
$\log[F_{\rm 40\mu m}/F_{\rm 20\mu m}]$, shown in color scale).  {\bf
  (a):} Each small square is a group of models for each set of $\mc$
and $\ms$. Inside each square, the four rows from top to bottom are
$\scl=0.1$, 0.32, 1, and 3.2~$\gcm$, and each column from left to
right is the inclination angle $\inc$ from an edge-on view to a
face-on view.  {\bf (b):} Similar to Panel (a), but calculated from
SEDs with a foreground extinction of $\av=100$.  {\bf (c):} Similar to
Panel (a), but each square shows the [40~$\mu$m]$-$[20~$\mu$m] color
averaged over $\scl$ and $\inc$ at each $\mc$ and $\ms$.  {\bf (d):}
Similar to Panel (c), but calculated from SEDs with a foreground
extinction of $\av=100$.  The dashed lines are where $\ms/\mc=0.01$,
0.03, 0.1, and 0.3.  The solid lines are where $\ms/\menv=0.01$, 0.1,
and 1. Here $\menv$ is averaged over $\scl$ for each $\mc$ and $\ms$.}
\label{fig:sed_mid}
\end{center}
\end{figure*}

\begin{figure*}
\begin{center}
\includegraphics[width=0.48\textwidth]{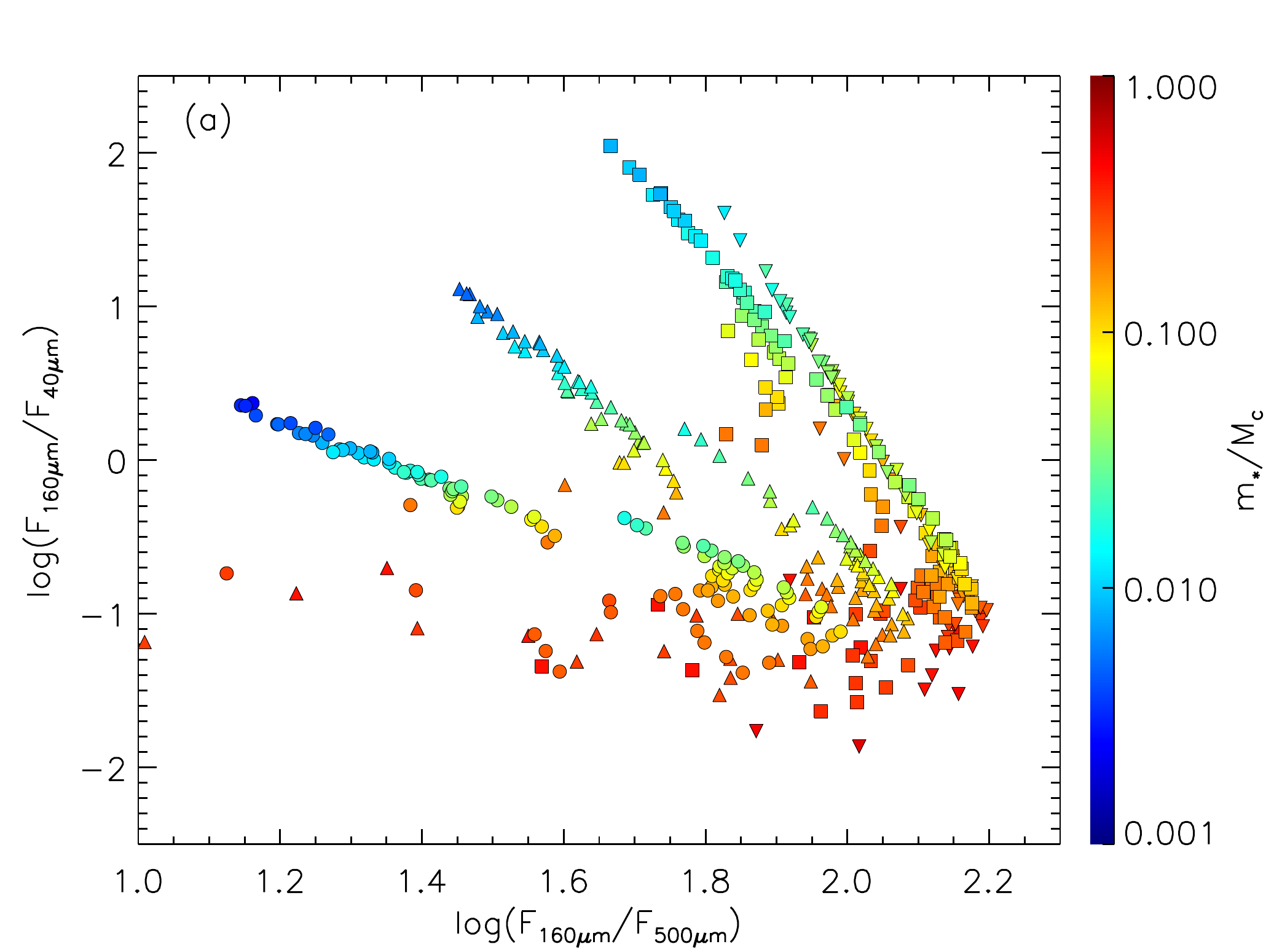}
\includegraphics[width=0.48\textwidth]{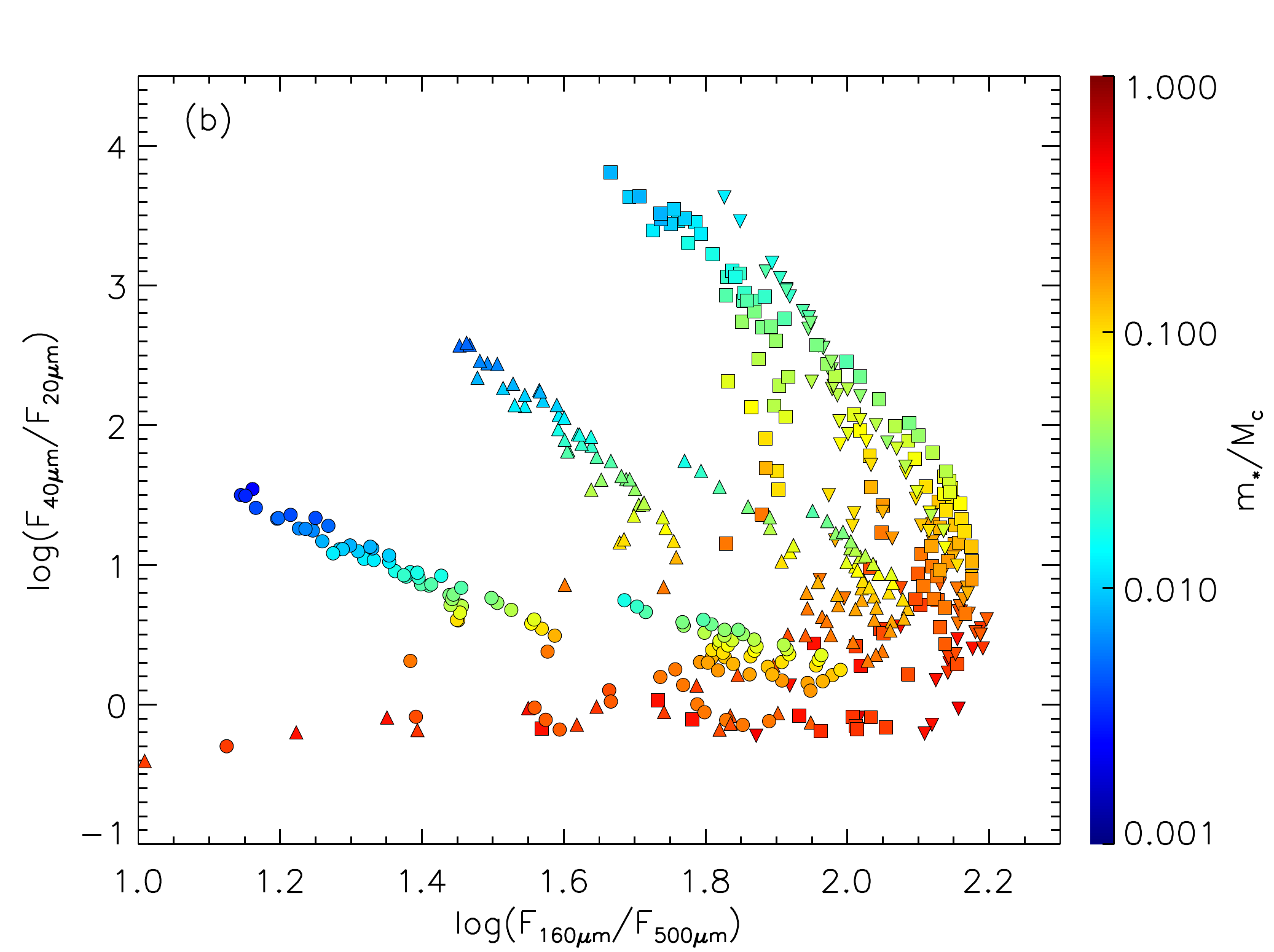}\\
\includegraphics[width=0.48\textwidth]{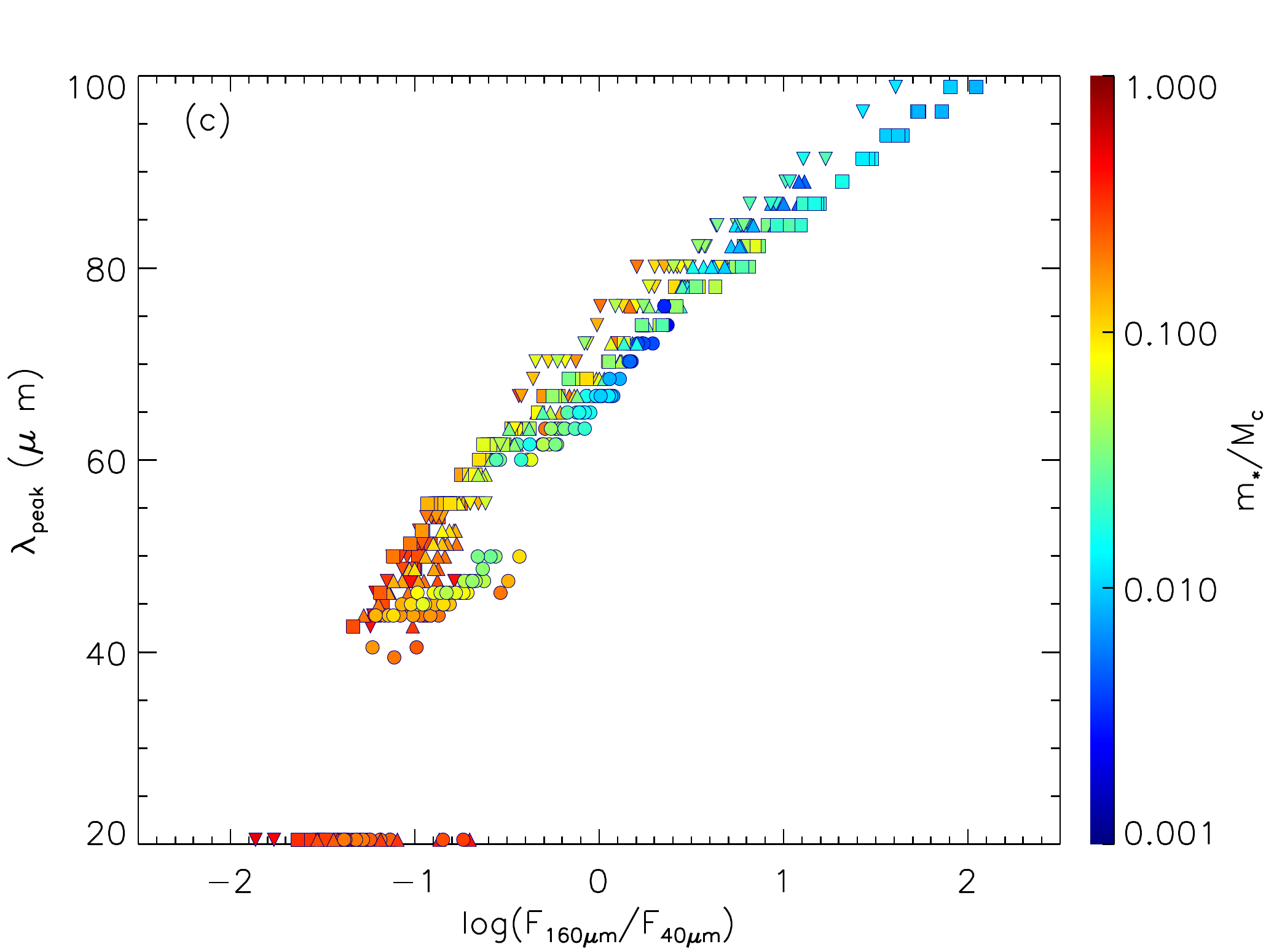}
\includegraphics[width=0.48\textwidth]{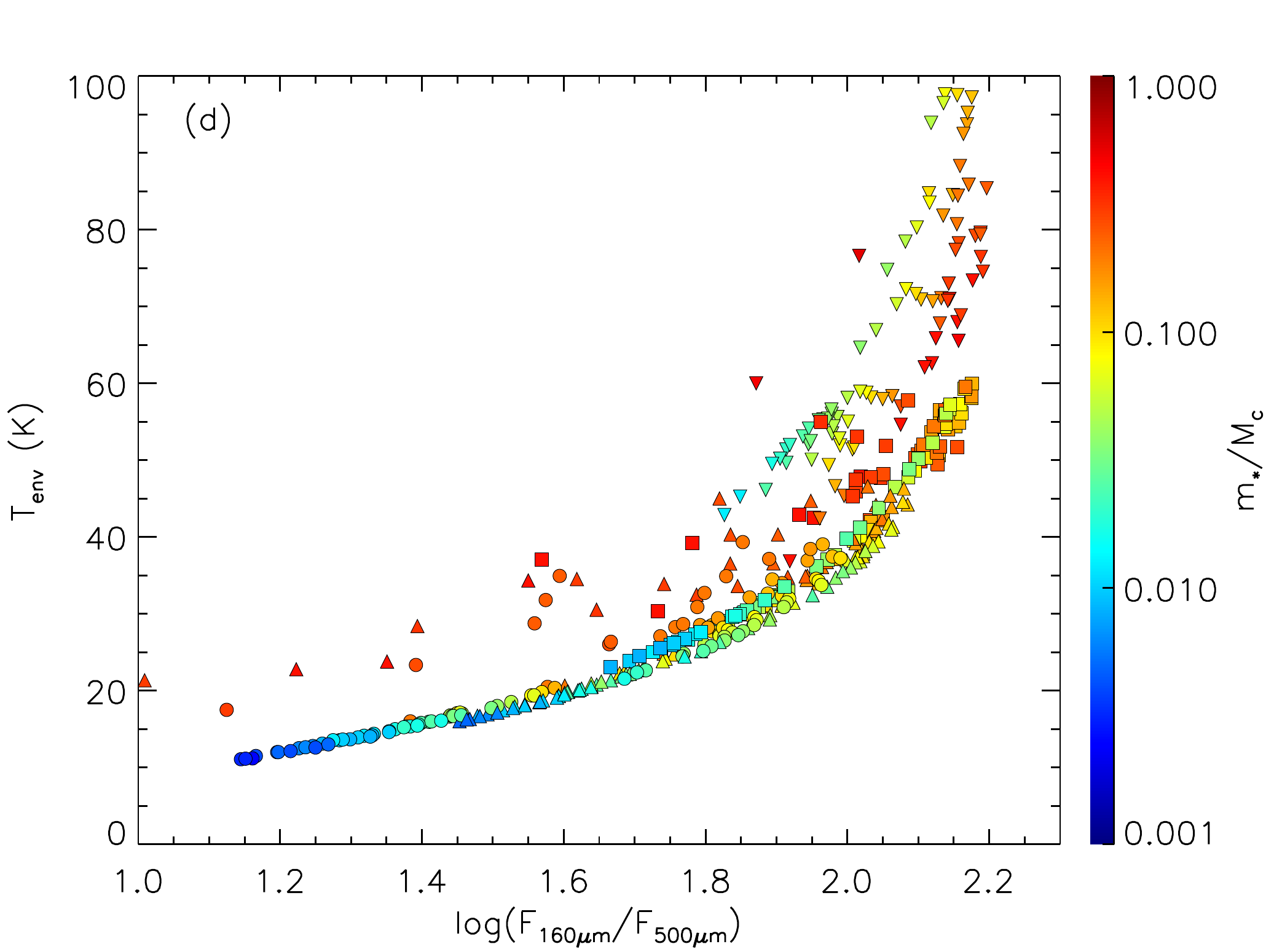}\\
\caption{Correlations between different SED characteristics in the model grid. 
{\bf (a):} the [160~$\mu$m]$-$[500~$\mu$m] color and the [160~$\mu$m]$-$[40~$\mu$m] color.
{\bf (b):} the [160~$\mu$m]$-$[500~$\mu$m] color and the [40~$\mu$m]$-$[20~$\mu$m] color.
{\bf (c):} the [160~$\mu$m]$-$[40~$\mu$m] color and the SED peak wavelength.
{\bf (d):} the [160~$\mu$m]$-$[500~$\mu$m] color and the mass-weighted mean temperature in the whole envelope.
The colors of the points show the evolutionary stages indicated by the parameter $\ms/\mc$,
from blue for early stage sources to red for later stage sources.
For models with the same $\mc$, $\scl$ and $\ms$, only the one with an inclination of $60^\circ$ is shown.
The circles, triangles, squares, upside-down triangles are models with $\scl=0.1$, 0.32, 1, and 3.2 $\gcm$,
respectively.}
\label{fig:sed_cor}
\end{center}
\end{figure*}

Figure \ref{fig:sed_fiducial} shows example SEDs from the model grid.
From top to bottom, the four panels show how the SED is affected by
the viewing inclination angle $\inc$, growth of the protostellar mass
$\ms$, the mass surface density of the star-forming environment $\scl$,
and the initial core mass $\mc$.

Panel (a) compares the SEDs of a same physical model
($\mc=60~M_\odot,~\scl=1~\gcm,~\ms=8~M_\odot$), but viewed at
different inclination angles.  From an edge-on view ($\cos\inc=0$) to a face-on view ($\cos\inc=1$),
the SED at shorter wavelengths increases, while the SED at wavelengths
longer than about 100~$\mu$m (i.e., longer than the wavelengths of the
SED peak) does not change.  
In this particular physical model ($\mc=60~M_\odot,~\scl=1~\gcm,~\ms=8~M_\odot$),
the opening
angle of the outflow cavity is $\thetaw\approx 24^\circ$
($\cos\thetaw\approx 0.9$).  The SED becomes flat when the viewing
inclination angle is smaller than this (i.e., when the line of sight
toward the protostar goes through the outflow cavity).

Panel (b) shows how the growth of the protostar affects the SED along
one evolutionary track ($\mc=60~M_\odot,~\scl=1~\gcm$).  A typical
viewing inclination angle of $\inc=61^\circ$ is used here.  As
the protostar grows, the fluxes at all wavelengths increase,
especially at the shorter wavelengths.  This is not only because of
the increase in total luminosity with the growth of protostar, but
also because of the gradual opening-up of the outflow cavity.  As the
flux is increasing, the SED peak also moves from about $100~\mu$m at
early stages to about $50~\mu$m at later stages. The SED becomes flat
at wavelengths $<70~\mu$m once the outflow cavity becomes wider than
the viewing inclination angle (the SED for $\ms=24~M_\odot$).

Panel (c) compares the SEDs of models with same $\mc$ and $\ms$, but
in environments with different $\scl$.  The mass surface density of
the environment affects the SED in several ways.  First, for these
four models, the luminosity is higher in a higher mass surface density
environment, which affects the height of the far-IR peak of the SED.
This is true in general, especially at earlier stages when the
accretion luminosity is dominant, but is affected by the detailed
evolutionary histories (see Panel (e) of Figure \ref{fig:history}).
Second, the mid-IR fluxes are mainly determined by the emission from
the accretion disk and also the dust inside of the outflow cavity,
after being extincted by the envelope. Since the envelope extinction
is lower in the low $\scl$ case, higher mid-IR fluxes are seen.
Third, the outflow cavity develops faster with the growth of the
protostar in the low $\scl$ case, which also makes the mid-IR fluxes
higher and the far-IR peaks at a shorter wavelengths.  Fourth,
different protostellar evolutionary tracks cause the input stellar
temperature to be significantly different among these mass surface
densities.  In addition, it is also worth noting that the depth of the
silicate feature around 10 $\mu$m is not monolithically dependent on
the mass surface density. It is deepest for an intermediate mass
surface density environment. This is because the silicate feature is
mostly caused by the optical depth in the disk, and in the high $\scl$
cases, the mid-IR fluxes from the disk start to be buried by the
emission from the warm envelope (see Paper II).

Panel (d) compares the SEDs of models with same $\scl$ and $\ms$, but
with different $\mc$.  With a more massive initial core, the protostar
is more embedded, leading to a higher contrast between the far-IR SED
and the shorter wavelength SED.  In the figure, SEDs with a foreground
extinction of $\av=100$, which corresponds to a mass surface density
of $0.3~\gcm$, are also shown. The SEDs at all wavelengths shorter
than about 100~$\mu$m are affected.

\subsubsection{Detailed Features of the SEDs}
\label{sec:sed_features}

In this section, we discuss several characteristics of the SEDs and
their distributions in the parameter space of the model grid.  These
characteristics include the wavelength of the SED peak, the bolometric
temperature, the far-IR slope at wavelengths around $160-500~\mu$m,
and the mid-IR slope at wavelengths around $20-40~\mu$m.  As the
results of the previous section have shown, these characteristics of
the SED are directly affected by the initial conditions, evolutionary
stages, and viewing inclination angles.  
While SED fitting works best with a fully sampled SED from MIR to FIR,
these characteristics may help to
constrain some of the conditions of massive protostars
in circumstances where fully sampled SEDs are not available.

As shown in Figure \ref{fig:sed_fiducial}, the wavelength of the SED
peak $\lambda_\mathrm{peak}$ in the far-IR is not very sensitive to
the viewing inclination or the foreground extinction (except for
face-on sources), and only sensitive to the physical conditions of the
source (in our case, determined by the initial conditions $\mc$,
$\scl$, and the protostellar mass $\ms$).  Figure \ref{fig:sed_peak}
shows the distributions of $\lambda_\mathrm{peak}$ in the parameter
space of the model grid.  The SED becomes flat and therefore
$\lambda_\mathrm{peak}\lesssim 30~\mu$m once the opening angle
is large and the line of sight toward the protostar goes through the
outflow cavity.  Such SEDs are also highly sensitive to the foreground
extinction.  Since foreground extinction is common in massive star
forming regions and quite uncertain, we expect to see the situation
represented in the right panels more often in real observations.  For
the models with viewing inclination angles larger than the opening
angle of the outflow cavity, the effect of the foreground extinction
on $\lambda_\mathrm{peak}$ is minor.  There is a weak dependence of
$\lambda_\mathrm{peak}$ on the mass surface density of the environment
$\scl$.  In early stages, the high $\scl$ models have SED peaks at
about $90~\mu$m, while the low $\scl$ models have SED peaks at about
$70~\mu$m.  In later stages, the high $\scl$ models have SED peaks at
about $70~\mu$m and the low $\scl$ models have SED peaks at about
$50~\mu$m, if the viewing inclination angle is larger than the opening
angle of the outflow cavity.  With a foreground extinction, the
difference is even smaller.  After averaging over inclination angles
and mass surface densities of the environment (lower panels), it is
evident that $\lambda_\mathrm{peak}$, with or without foreground
extinction, is dependent on the evolutionary stages indicated by
$\ms/\mc$ or $\ms/\menv$, especially when the protostellar mass
$\ms\gtrsim 8~M_\odot$.  At the stage of $\ms/\menv=1$, which marks
the transition from the main accretion phase to the envelope clear-up
phase, $\lambda_\mathrm{peak}$ is about $50-60~\mu$m in the case of
foreground extinction $\av=100$.  At the stage that $\ms/\menv=0.1$,
$\lambda_\mathrm{peak}$ is about $60-70~\mu$m.

Figure \ref{fig:sed_midtofar} shows how the [160~$\mu$m]$-$[40~$\mu$m]
color (defined as $\log[F_{160\mu\mathrm{m}}/F_{40\mu\mathrm{m}}]$)
depends on the initial conditions, protostellar mass, and viewing
inclination angle. The [160~$\mu$m]$-$[40~$\mu$m] color is related to
$\lambda_\mathrm{peak}$, which is between these two wavelengths.  In
fact, the [160~$\mu$m]$-$[40~$\mu$m] color is easier to determine,
since it does not require a well sampled SED around the peak position,
although it is more affected by the inclination or the foreground
extinction since the flux at $40~\mu$m is used.  However, as Figure
\ref{fig:sed_midtofar} shows, the effects of inclination and possible
foreground extinction on the [160~$\mu$m]$-$[40~$\mu$m] color are
modest. The value of $\log[F_{160\mu\mathrm{m}}/F_{40\mu\mathrm{m}}]$ increases by
about 1 when varying the inclination from a face-on view to an edge-on
view, and increases by about 0.3 with a foreground extinction of
$\av=100$. Overall, it changes by up to about 4 over the whole
parameter space.  The dependence of the [160~$\mu$m]$-$[40~$\mu$m]
color on the mass surface density of environment $\scl$ is also
relatively weak (the increase is by $\lesssim 1$ from the high mass
surface density case to the low mass surface density case).  It is
sensitive to the evolutionary stage, indicated by $\ms/\mc$ or
$\ms/\menv$.  Therefore the [160~$\mu$m]$-$[40~$\mu$m] color may be
used as an indicator of the evolutionary stage of massive protostars,
in addition to the peak position of the SED.  The SEDs at the stage
when $\ms/\menv=1$ have
$F_{160\mu\mathrm{m}}/F_{40\mu\mathrm{m}}\approx -1$ and those at the
stage of $\ms/\menv=0.1$ have
$F_{160\mu\mathrm{m}}/F_{40\mu\mathrm{m}}\approx 0$.

Besides the peak wavelength, one can also use the flux-weighted mean
wavelength as an indicator of the evolutionary stage. The concept that
is related to this is the bolometric temperature, which is defined as
the temperature of a blackbody having the same mean frequency as the
observed SED (\citealt[]{Myers93}), which can be written as
$\tbol\equiv1.25\times 10^{-11}\langle\nu\rangle~\mathrm{K~Hz}^{-1}$.
Here, for $\langle\nu\rangle$, we use the flux-weighted mean frequency
$\int\nu{F_\nu}d\nu/\int{F_\nu}d\nu$ in the wavelength range of
$\lambda> 1~\mu$m.  This is because often only infrared data are
available when constructing SEDs for massive protostars, and the
fluxes at shorter wavelengths normally suffer from high levels of
extinction, since massive protostars are typically highly embedded in
high mass surface density clouds.  Compared to the peak wavelength of
the SED, the bolometric temperature is not so affected by whether the
SED is well sampled around the peak. However, it is sensitive to the
inclination and the foreground extinction, since the short wavelength
fluxes are used to determine $\tbol$. Therefore one should be cautious
when using the bolometric temperature to estimate the evolutionary
stage, especially for individual sources. For a sample with a large
number of sources, we expect the effects of different viewing
inclination angles will be averaged out.

Figure \ref{fig:tbol} shows how the bolometric temperatures are
affected by the initial conditions, the evolutionary stages, the
viewing inclination angles, and foreground extinction.
From Panel (a) and (b), we can see that $\tbol$ is distinctively
different for the inclinations at which the line-of-sight toward the
central protostar goes through the outflow cavity, compared to those
at which the line-of-sight goes through the envelope.  In the former
case (low inclination angles), $\tbol$ is $\gtrsim 300$~K without
foreground extinction, and about 200~K if a foreground extinction of
$\av=100$ is applied.  In the latter case (high inclination angles),
$\tbol$ is less affected by the foreground extinction.  $\tbol$ ranges
from about 30~K to about 120~K without foreground extinction, and up
to about 80~K with a foreground extinction of $\av=100$.  $\tbol$ is
slightly dependent on the mass surface density of the environment
$\scl$, which makes $\tbol$ increase by about 30~K from the highest
$\scl=3.2~\gcm$ to the lowest $\scl=0.1~\gcm$ in the case without any
foreground extinction, which will lower the effect.  At a typical
inclination of $60^\circ$ and averaging over the mass surface density
of the environment, the bolometric temperature is sensitive to the
evolutionary stage indicated by $\ms/\mc$ or $\ms/\menv$, i.e., how
embedded the protostar is, as Panels (c) and (d) show.  At the stage
where $\ms/\menv=1$ (which corresponds to the transition from Class 0
to I in low-mass star formation), $\tbol\approx 200 - 300$~K in the
case of no foreground extinction, and $\tbol\approx 100$~K with a
foreground extinction of $\av=100$.  These values are higher than
$\tbol=70$~K, which is commonly used as the boundary between the Class
0 and Class I sources in low-mass star formation studies
(\citealt[]{Chen95}).  However, in the cases of low $\scl$ and low
$\mc$, which is closer to the situation of normal low-mass star
formation, this transition does occur at about $\tbol=70$~K.

Figure \ref{fig:sed_far} shows the far-IR/sub-mm slope of the SED
defined by $\log(F_{160\mu\mathrm{m}}/F_{500\mu\mathrm{m}})$ in the
parameter space.  As Figure \ref{fig:sed_fiducial} has shown, the
slope in this wavelength range is not affected by the inclination or
foreground extinction, therefore we only consider its dependence on
$\mc$, $\scl$ and $\ms$.  As Figure \ref{fig:sed_far} shows, the
far-IR slope of the SED has a clear dependence on the protostellar
mass. As the protostar grows, the far-IR slope becomes steeper, which
can be also seen in Panel (b) of Figure \ref{fig:sed_fiducial}.
Unlike the peak wavelength or colors of other wavelength ranges
discussed above, the far-IR slope has only a weak dependence on the
initial core mass (or the current envelope mass), especially when the
protostellar mass $\ms \gtrsim 10~M_\odot$.  The mass surface density
of the environment also affects the far-IR slope of the SED.  In a
high mass surface density environment, the higher column density of
the envelope causes more shorter wavelength emission from the inner
hot regions, such as disk or innermost envelope, to be shifted to
longer wavelengths, leading to a steeper slope of the far-IR/sub-mm
SED.  Although the far-IR slope of the SED is not so affected by the
viewing inclination angle or the foreground extinction, and thus is
mostly only dependent on the physical conditions of the source, in
real observations, it is affected by the ambient clump material, which
is not included in our models. Also the exact value of the slope is
affected by dust emissivity properties.

Figure \ref{fig:sed_mid} shows the slope of the SED from 20 to 40
$\mu$m in the parameter space.  The SED slope at this wavelength range
can be significantly affected by the mass surface density of the
environment, $\scl$, from a relatively flat slope with
$\log(F_{40\mu\mathrm{m}}/F_{20\mu\mathrm{m}})\approx 0 - 1$ to a
steep slope with $\log(F_{40\mu\mathrm{m}}/F_{20\mu\mathrm{m}})\approx
3 - 4$, given with the same initial core mass and protostellar mass.
This is caused by the significant effects of the column density of the
envelope on the fluxes around $10-20~\mu$m which is mainly from the
disk, dust inside of the outflow cavity, or the innermost hot
envelope.  After averaging over $\scl$, one can clearly see that the
SED slope at 20 to 40 $\mu$m depends on the evolutionary stages
indicated by $\ms/\menv$ or $\ms/\mc$, i.e., how embedded the
protostar is, rather than simply on $\ms$, i.e., the growth of the
protostar.  The SED slope at this wavelength range is also affected by
the viewing inclination angle, but normally within a range of about
one order of magnitude, not considering the extreme cases such as a
face-on source.  The foreground extinction has a modest effect on the
slope
compared with that due to the evolution, 
changing $\log(F_{40\mu\mathrm{m}}/F_{20\mu\mathrm{m}})$ by
about 0.6 for a foreground extinction of $\av=100$.

Figure \ref{fig:sed_cor} shows how the SED features discussed in this
section, including the [160~$\mu$m]$-$[500~$\mu$m] color,
[160~$\mu$m]$-$[40~$\mu$m] color, [40~$\mu$m]$-$[20~$\mu$m] color, and
the SED peak wavelength correlate to each other.  As discussed above,
all these colors at different wavelength ranges can be used to
estimate the evolutionary stages, but they are sensitive to different
components of the source.  The [160~$\mu$m]$-$[500~$\mu$m] color is
mostly affected by the envelope, while the [160~$\mu$m]$-$[40~$\mu$m]
or [40~$\mu$m]$-$[20~$\mu$m] colors are more sensitive to the inner
warmer regions, including the outflow cavity wall.  Panel (a) and (b)
clearly show that the long-wavelength color and the short wavelength
colors correlate to each other, but the correlations are affected by
the initial condition of the source, especially the environmental mass
surface density $\scl$.  Panel (c) shows that the
[160~$\mu$m]$-$[40~$\mu$m] color is well correlated with the peak
wavelength of the SED, except for some models at late stages or very
wide outflow opening angles, for which the SED peaks are at
wavelengths $\lesssim 20~\mu$m. As mentioned above, both of these two
features can be used to indicate the evolutionary stages, but the
[160~$\mu$m]$-$[40~$\mu$m] color is easier to determine since it does
not require a well sampled SED around its peak position.

\subsubsection{Flashlight Effect}
\label{sec:flashlight}

\begin{figure}
\begin{center}
\includegraphics[width=0.5\columnwidth]{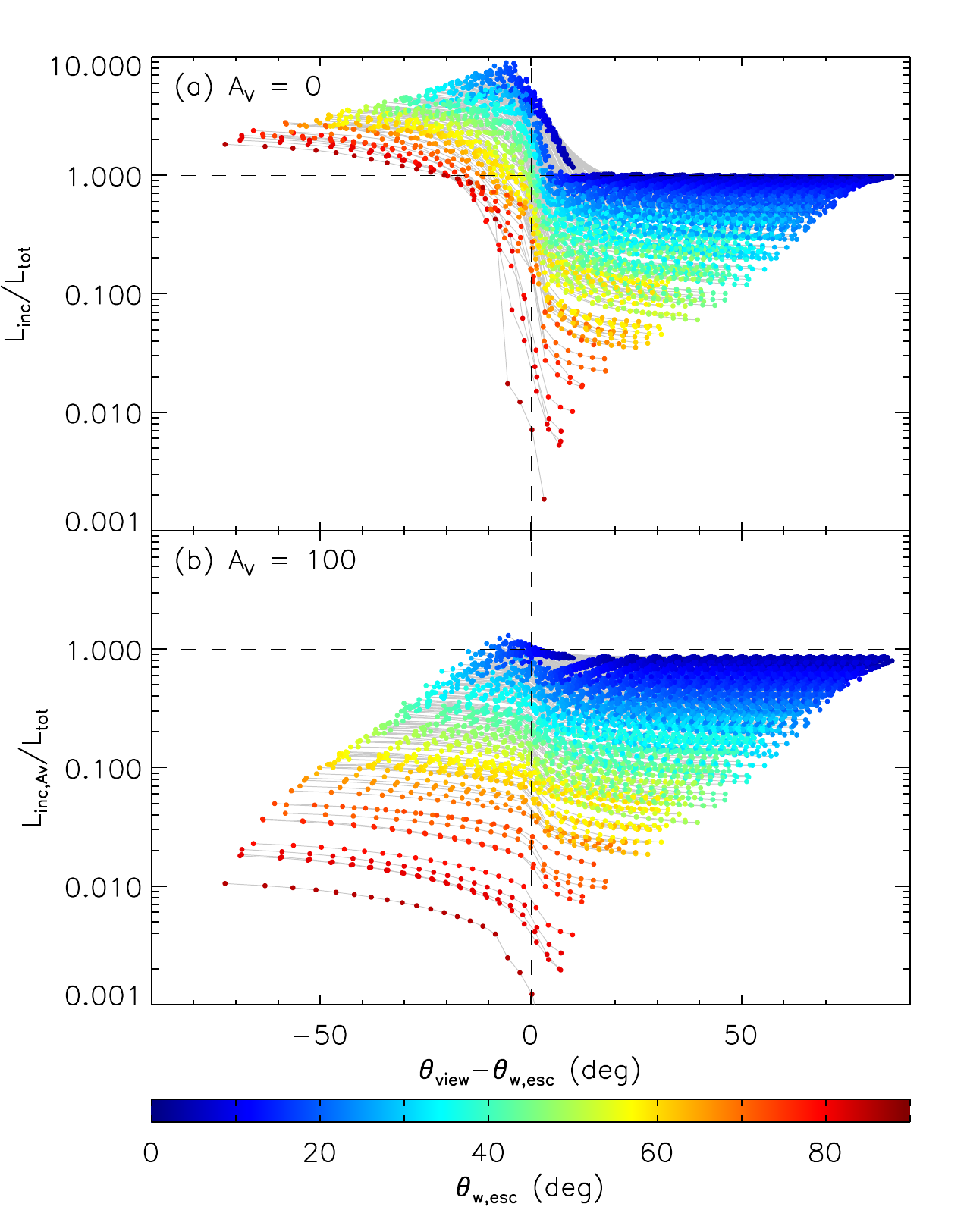}
\caption{{\bf (a):} Change of the ratio between $\linc$, the bolometric
luminosities inferred by assuming isotropic radiation (i.e., without
correction for the viewing inclination angle), and $\ltot$, the true
bolometric luminosities, with the viewing angle relative to the
opening angle of the outflow cavities ($\inc-\thetaw$).  The lines
connect the models with same values of $\mc, \scl, \ms$, but different
$\inc$.  The color indicates the opening angle of the outflow
cavities, $\thetaw$, in each model.  {\bf (b):} Similar to Panel (a),
but with a foreground extinction of $\av=100$ applied.  The $y$-axis
is now the ratio between the bolometric luminosity inferred from the
extincted SED ($L_{\mathrm{inc},A_V}$) and the true bolometric
luminosity of the source.}
\label{fig:flashlight}
\end{center}
\end{figure}

As discussed above, the observed SED is highly dependent on the
viewing inclination angle, because the existence of the low-density
outflow cavity allows more radiation to escape from the polar
direction, which is known as the ``flashlight effect''.  This causes
the bolometric luminosity integrated from the observed SED to deviate
from the true bolometric luminosity of the source by a factor which
depends on the inclination.  Panel (a) of Figure \ref{fig:flashlight}
shows the ratios between the bolometric luminosities inferred from
SEDs by assuming isotropic radiation, $\linc$, and the true bolometric
luminosities, $\ltot$, in the model grid.  As the figure shows, the
degree of the flashlight effect is almost completely dependent on two
factors: the opening angle of the outflow cavity, $\thetaw$, and the
viewing inclination angle, $\inc$.  In most of the cases where
$\inc<\thetaw$, $\linc$ will overestimate $\ltot$ by a factor up to
10, while in most of the cases where $\inc>\thetaw$, $\linc$ will
underestimate the true luminosity $\ltot$.  For an opening angle of
about $50^\circ$, the inferred luminosity $\linc$ overestimates the
true luminosity by a factor of $2-3$ for a face-on view and
underestimates the true luminosity by a factor of 10 for an edge-on
view. For an opening angle of about $20^\circ$ or smaller, the
inferred luminosity is close to the true luminosity for most
inclinations, except at a face-on view.

In real observations, the luminosity directly inferred from the SED
further deviates from the true luminosity due to possible foreground
extinction.  Panel (b) of Figure \ref{fig:flashlight} shows the ratios
between the luminosities inferred from SEDs with $\av=100$ and the
true bolometric luminosities of the sources.  In such a case, most of
the inferred luminosities are underestimating the true luminosities.
Since the foreground extinction mainly lowers the SEDs at shorter
wavelengths, which mainly affects the SEDs at lower viewing
inclination angles, the flashlight effect is not so much affected by
the foreground extinction, if the inclination angle is larger than the
opening angle of the outflow cavity.  
In this case, the flashlight effect is significantly affected by the opening
angle of the outflow cavity, but only very weakly affected by the inclination angle.
With a foreground extinction of
$\av=100$, for an opening angle of about $50^\circ$, the inferred
luminosity underestimates the true luminosity by a factor of about 10,
no matter the viewing inclination angle.


\subsubsection{Temperature Evolution in the Envelope}


\begin{figure}
\begin{center}
\includegraphics[width=0.5\columnwidth]{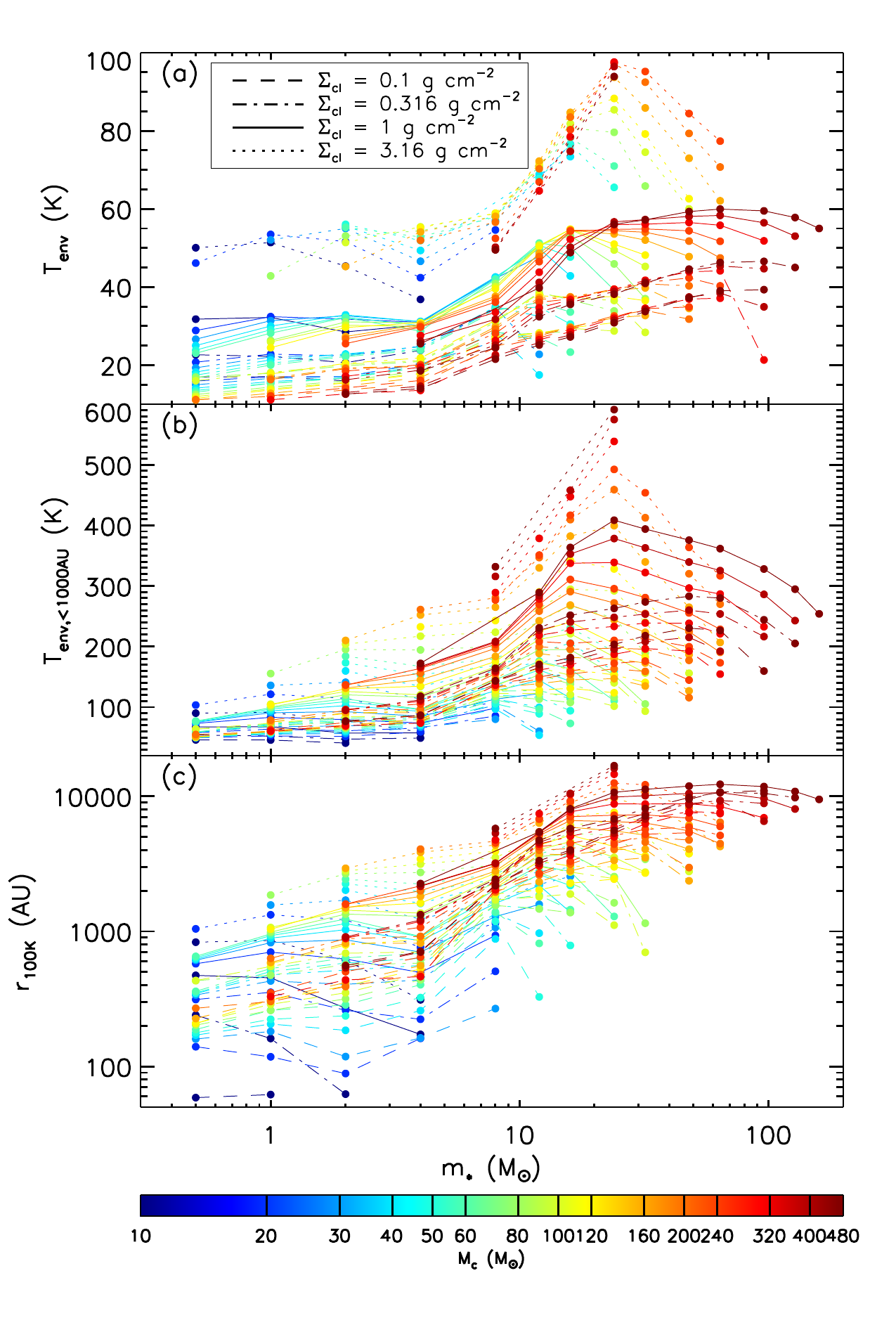}
\caption{{\bf (a):} The evolutions of the mass-weighted mean temperature in the
whole envelope with the growth of protostar along each evolutionary
track.  Different initial conditions of the evolutionary tracks are
shown by different colors ($\mc$) and line styles ($\scl$).  {\bf
  (b):} Similar to Panel (a), but showing the evolutions of the
mass-weighted mean temperature in the envelope within 1000~AU from the
protostar.  {\bf (c):} The evolutions of the size of the part of
envelope which has a mass-weighted mean temperature of 100~K.}
\label{fig:tevo}
\end{center}
\end{figure}

\begin{figure}
\begin{center}
\includegraphics[width=0.5\columnwidth]{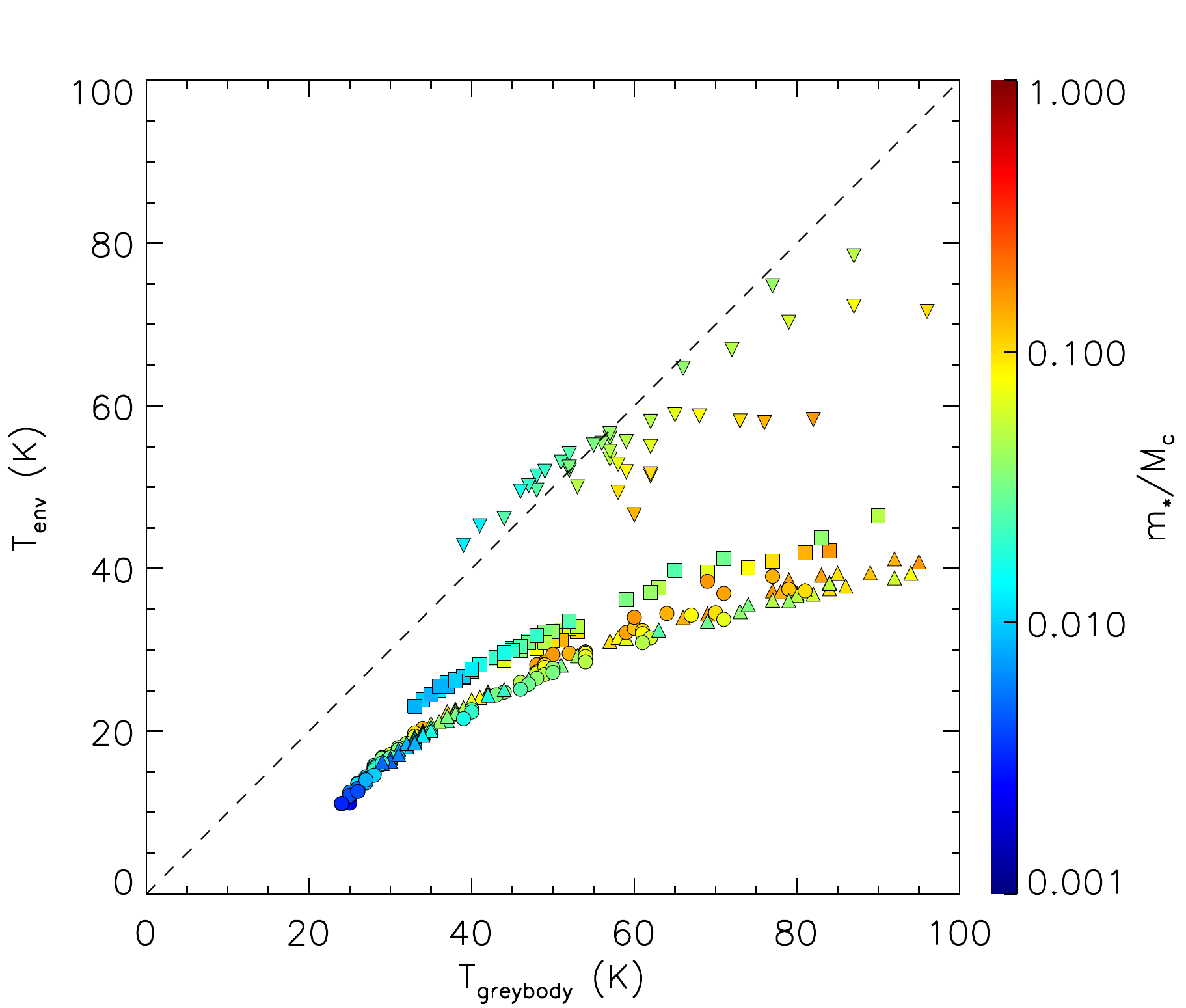}
\caption{
The mean mass-weighted temperature in the envelope, $T_\mathrm{env}$,
compared with the temperature obtained by grey-body fitting to the
fluxes at 100, 160, 250, 350, 500 and 850 $\mu$m,
$T_\mathrm{greybody}$.  The color shows the evolutionary stages
indicated by the parameter $\ms/\mc$, from blue for early stage
sources to red for later stage sources.  For models with the same $\mc$,
$\scl$ and $\ms$, only the one with an inclination of $60^\circ$ is
shown.  The circles, triangles, squares, upside-down triangles are
models with $\scl=0.1$, 0.32, 1, and 3.2 $\gcm$, respectively.
The dashed line indicates $T_\mathrm{env}=T_\mathrm{greybody}$.}
\label{fig:tgrey}
\end{center}
\end{figure}

The radiative transfer simulation also predicts the dust temperature
profiles for the models in the grid.  Panel (a) of Figure
\ref{fig:tevo} shows the evolutions of the mass-weighted mean
temperature in the whole envelope as a function of protostellar mass
under different initial conditions.  The envelope temperature has a
clear dependence on the protostellar mass $\ms$ and the mass surface
density of the star-forming environment, $\scl$, but only a weak
dependence on the initial core mass, $\mc$.  In the low mass surface
density cases ($\scl=0.1$ and $0.3~\gcm$), at early stages, the core
has a mean temperature of $\lesssim 20$~K, while with $\scl=1~\gcm$,
the mean temperature is about 30~K, and in the high mass surface
density case, the temperature reaches about 50~K, even in the earliest
stages.  As the protostar grows, the envelope becomes warmer. There is
a significant increase in the envelope temperature around
$\ms=4~M_\odot$. The peak temperature is reached at $\ms\gtrsim
20~M_\odot$.  The peak temperature of the high mass surface density
core is about 100~K.  The mass surface density of the star-forming
environment affects the temperature in the envelope in several ways.
As discussed in \S\ref{sec:model}, in a high mass surface density
environment, the core is more compact and collapses with a higher
accretion rate, leading to a higher luminosity. The high luminosity
and the small size of the core combined make the temperature higher in
such a core.  In some models, the temperature starts to decrease in
the final stages. This is because of the wide opening angle of the
outflow cavity starts to suppress the heating from the protostar due
to escape of radiation through the low-density outflow cavity.

The different luminosities and envelope densities in the various
$\scl$ environments have a more significant impact on the temperature
in the innermost region of the envelope.  Panel (b) shows the
mass-weighted mean temperature in the region of the envelope within
1000~AU from the protostar.  On this scale, the peak temperature of
the high mass surface density core can reach about $400-600$~K at
around $\ms=20~M_\odot$.  For most of the models in low mass surface
density environments ($\scl=0.1$ or $0.3~\gcm$), the mean envelope
temperature on the 1000~AU scale is below 100~K in early stages
($\ms\lesssim4~M_\odot$) and below about 200~K in later stages.

Hot core chemistry is initiated when the temperature reaches about
100~K, at which point dust grain ice mantles are largely sublimated
and various complex molecules are released to the gas phase, producing
important observational diagnostics for the protostellar stage of
massive star formation.  Panel (c) of Figure \ref{fig:tevo} shows the
evolution of the hot core size (defined as the size of the envelope
that has a mass-weighted mean temperature of 100~K) with the growth of
the protostar under different initial conditions.  In the early
stages, the hot core region is only present within a small zone of
several hundreds of AU. Later on, when $\ms\gtrsim 10~M_\odot$, the
hot core is typically of a scale of several thousands of AU.  In
addition, the size of the hot core has clear dependence on the mass
surface density of the environment in the early stages. However, this
dependence becomes weaker in the later stages, when the hot core is
fully developed.  The dependence of the envelope temperature on the
mass surface density of the environment suggests that even for
protostellar cores with the same mass and at the same evolutionary
stages, the chemistry can be significantly affected by the
star-forming environment, which is not only true for massive
protostars, but also for the low-mass protostars forming alongside
them (\citealt[]{ZT15}).

For prestellar core or early-stage protostellar cores, grey-body
fitting is often performed to estimate the temperature of the cores
(e.g., \citealt[]{Elia17}).  Figure \ref{fig:tgrey} shows how the mean
mass-weighted temperature in the envelope differs from the grey-body
fitted temperature in our model grid.  Here the grey-body temperature is
obtained by fitting the model fluxes at 100, 160, 250, 350, 500, and
850 $\mu$m following the greybody spectrum $\fnu\propto\nu^\beta
B_\nu(T)$, where the dust emissivity index $\beta$ is set to be 2 and
$B_\nu$ is the Planck function.  The fitting gives higher
$T_\mathrm{greybody}$ for later-stage protostellar sources, which are
not shown in this figure.  For the models with $\scl =0.1-1~\gcm$, the
mean temperature in the envelope is always $\lesssim 40\:$K, as shown
above. The greybody temperature is always higher than $T_\mathrm{env}$
and as the source evolves, it soon increases from about 20 K to
$\gtrsim$100 K.  In the case of $\scl=3.2~\gcm$, the greybody
temperature follows $T_\mathrm{env}$ better in the early stages.  We
note that our model grid only covers the protostellar phase of massive
star formation when a protostar $\gtrsim 1~M_\odot$ has formed, while
the grey-body fit works best for the prestellar phase or early
protostellar phase.

Panel (d) of Figure \ref{fig:sed_cor} shows how the mean temperature
in the envelope correlates with the [160~$\mu$m]$-$[500~$\mu$m] color.
These two properties are well correlated, especially for models with
$\scl=0.1-3~\gcm$ and at evolutionary stages with
$\ms/\mc\lesssim0.1$.  This correlation, taking into account the
correlation between $T_\mathrm{env}$ and $T_\mathrm{greybody}$, is
consistent with that found by \citet[]{Elia17} from a large sample of
cores/clumps observed in the Hi-GAL survey.  Their results extend
this correlation further to even earlier stages with $T<10$ K, which is
not covered by our model grid.

\subsection{Caveats of the Model Grid}
\label{sec:caveat}

Here we describe several important caveats and limitations of our SED
model grid.
First, the model grid is based on the Turbulent Core model of
\citet{MT03}, in which massive cores are embedded in a larger ambient
clump that will form the star cluster.  The cores are assumed to be in
pressure equilibrium with the ambient clump, and the surface pressure
on the core is characterized using the mean mass surface density of
the clump $\scl$, which is why $\scl$ is used as one of the primary
parameters of the model grid. However, except via this parameter, the
clump itself is not included in the radiative transfer simulations,
which is not realistic. We ignore the ambient material because it
would bring in more free parameters about the sizes and density
distributions of the clumps, which are highly uncertain.  The ambient
clump may affect the SEDs in two ways. First, it provides additional
foreground extinction to lower the fluxes at shorter
wavelengths. Second, it provides additional emission at longer
wavelengths. The former effect can be compensated by the free
parameter of foreground extinction $\av$, but the latter effect is not
taken into account in our model grid.  Therefore, an aperture should
be carefully chosen at wavelengths $\gtrsim 70~\mu$m to exclude the
contribution of the ambient clump material in deriving the fluxes at
these wavelengths before performing the SED fitting, i.e., the model
grid fitting should be done on clump-envelope-background-subtracted
SEDs.

Second, in the current model grid, when constructing the SEDs, we have
summed up all the photons emitted in a specified direction, no matter
from where they emerge. That is to say, we are not applying any
aperture to exclude the emission from any part of the model core,
i.e., these are total SEDs, including from parts of the outflow that
extend beyond the core. 
However, observations of real sources at different wavelengths by
different instruments may be on different scales, which also need to
be considered when performing the SED fitting.

Third, the observed short wavelength fluxes at $\lesssim 8~\mu$m are affected
by PAH emission and thermal emission from very small grains that are
transiently heated by single photons, and these effects have not been
included in our radiative transfer models. Therefore we expect the
models are under-predicting the real fluxes at these wavelengths.  In
the example we show in \S\ref{sec:example}, we set the observed fluxes
at $\lesssim 8~\mu$m to be upper limits. However, users can freely
adjust which data points should be used as limits according to their
situation and needs, and the fitting program provides high flexibility for
setting the upper/lower limits and uncertainties (see
\S\ref{sec:fitting}).

Finally, some detailed features of the SEDs, such as the peak
wavelength and the long wavelength spectral index, may be affected by
the particular dust models used in the radiative transfer
simulations. Although the general trends of these features with the
initial/environmental conditions ($\mc, \scl$) and evolution ($\ms$)
discussed above should not change, the exact values may be affected.

\section{SED Fitting}
\label{sec:fitting}

We use $\chisq$ minimization to find the best model to fit the
observed SED. Assuming that we have observed flux densities $\fnuobs$
with upper and lower uncertainties of $\sigmau(\fnuobs)$ and
$\sigmal(\fnuobs)$ at wavelengths $\lambda_1, ..., \lambda_N$, and for
each model (i.e., each set of $M_c$, $\scl$, $m_*$, $\inc$, $d$,
$\av$) we have model flux densities $\fnumodav$ (see Equation \ref{eq:fnumodav}) at these wavelengths,
the reduced $\chisq$ is defined as
\begin{eqnarray}
\chisq & = & \frac{1}{N_\mathrm{total}}\left\{\sum_{\fnumodav>\fnufit}\left[\frac{\log\fnumodav-\log\fnufit}
{\sigmau(\log\fnufit)}\right]^2\right.\nonumber\\
 & + & \left.\sum_{\fnumodav<\fnufit}\left[\frac{\log\fnumodav-\log\fnufit} 
{\sigmal(\log\fnufit)}\right]^2\right\},\label{eq:chisq}
\end{eqnarray}
where $\fnufit$, $\sigmau(\log\fnufit)$ and $\sigmal(\log\fnufit)$ are
derived from $\fnuobs$, $\sigmau(\fnuobs)$, and $\sigmal(\fnuobs)$
(see \S\ref{sec:error}).  For $\fnuobs$ used as upper limits,
$\sigmal=\infty$, i.e., no contribution to the $\chisq$ if
$\fnumodav<\fnufit$, and for $\fnuobs$ used as lower limits,
$\sigmau=\infty$, i.e., no contribution to the $\chisq$ if
$\fnumodav>\fnufit$.  The total number of data points $N_\mathrm{total}$ contains
both normal data points and upper/lower limits.

During each fitting, we first search for a minimum $\chisq$ by varying
the foreground extinction $\av$ for each set of $(\mc,~\scl,~\ms,
\inc, d)$. We then compare these minimum $\chisq$ values to find the
best models in the 5 dimensional parameter space formed by different
$(\mc,~\scl,~\ms,~\inc,~d)$ (4 dimensions if an exact source distance
$d$ is provided).  We also select another group of best models. For
each set of $(\mc, \scl, \ms)$, we search for the minimum $\chisq$ by
varying $\av$, $d$, and $\inc$, and then compare these minimum
$\chisq$ values to find the best models in the 3 dimensional parameter
space formed by different $(\mc, \scl, \ms)$.  Therefore, each member
in this group of best models is a different physical model.  In the
released code, both groups are output and users can choose which to
use according to their need.

While the $\chisq$ defined in Equation \ref{eq:chisq} is used in the ranking
and selection of the best fitted models, we also define
\begin{equation}
\chisq_\mathrm{nonlimit}\equiv\chisq\frac{N_\mathrm{total}}{N_\mathrm{nonlimit}},
\end{equation}
where $N_\mathrm{nonlimit}$ is the number of data points that have
non-zero contributions to $\chisq$.  Note that for the same observed
SED, $N_\mathrm{nonlimit}$ is dependent on the model SEDs.  For
  example, for a data point used as an upper limit, if the model SED
  is higher than that data point, it is counted in
  $N_\mathrm{nonlimit}$, while if the model SED is lower than that
  data point, it is not counted in $N_\mathrm{nonlimit}$, since it is
  not contributing to $\chisq$.  Therefore $\chisq_\mathrm{nonlimit}$
  is the average deviation of the model SED from the constraining data
  points.

The SED fitting tool as well as the model grid are available for
  download.
\footnote{See
  \dataset[https://doi.org/10.5281/zenodo.1045606]{https://doi.org/10.5281/zenodo.1045606}.}
Currently, the fitting tool is written in IDL, therefore the IDL
software needs to be installed before running the fitting
program. However, knowledge of the IDL language is not necessary.  An
instruction file including detailed information about the structure of
the package, its installation, editing input parameters, running the
fitting program, and output files and figures are included in the
package.  In Section \ref{sec:example}, we will show an example of the
SED fitting (the same example is also included in the package) and
discuss the output results and figures.  Currently, the fitting
program is designed to fit SEDs of individual sources,
However, it can be easily adapted to perform recursive fitting to a
sample of SEDs.
It takes about 1 minute to fit an SED running the program on the processor of a typical current laptop or desktop computer.

\subsection{Treatment of the Errors}
\label{sec:error}

The fitting is performed in logarithmic space since the fluxes are
nonlinear with wavelength and most of the errors are best described as
certain percentages of the fluxes. 
Assuming the observed flux and its error at a certain wavelength are 
$\fnuobs$ and $\sigma(\fnuobs)$, 
i.e., $\fnuobs\equiv<\fnu>$ and $\sigma(\fnuobs)\equiv\mathrm{var}(\fnu)$,
the expectation and variance of
the fluxes in logarithmic space are related to these values by
\begin{eqnarray}
\log\fnufit \equiv <\log\fnu> & = & \log<\fnu>-\frac{1}{2\ln 10}\left[\frac{\mathrm{var}(\fnu)}{<\fnu>}\right]^2+...\nonumber\\
& = & \log\fnuobs-\frac{1}{2\ln 10}\left[\frac{\sigma(\fnuobs)}{\fnuobs}\right]^2+...~,
\end{eqnarray}
and
\begin{eqnarray}
\sigmal(\log\fnufit) = \sigmau(\log\fnufit) \equiv \mathrm{var}(\log \fnu) & = & \frac{1}{\ln 10}\frac{\mathrm{var}(\fnu)}{<\fnu>}+...\nonumber\\
& = &  \frac{1}{\ln 10}\frac{\sigma(\fnuobs)}{\fnuobs}+...~.
\end{eqnarray}
This is only valid when the percentage errors are small, since it is a
first-order approximation.  In real observations, due to the
uncertainties in brightness calibration, background subtraction, and
selection of apertures to integrate the emission, the percentage
uncertainties in the observed flux can easily be several $\times 10\%$
or higher, in which case the error becomes asymmetric around the
observed flux in the logarithmic space. Therefore we define the
following fluxes and errors in logarithmic space:
\begin{equation}
\log\fnufit \equiv \log\fnuobs,
\end{equation}
\begin{equation}
\sigmau(\log\fnufit) \equiv \log\left[1+\frac{\sigmau(\fnuobs)}{\fnuobs}\right],
\end{equation}
\begin{equation}
\sigmal(\log\fnufit) \equiv -\log\left[1-\frac{\sigmal(\fnuobs)}{\fnuobs}\right],
\end{equation}
which are simply obtained by converting the data points and error bars to
logarithmic space.  The log space flux densities and errors in
these two methods start to differ significantly when the percentage
error becomes $\gtrsim 50\%$.  We allow both types of conversion in
the fitting program. If the first method is used, the input upper and
lower errors need to be same, while if the second method is used, the
users can input different upper and lower errors for each flux. We use
the second method in the example discussed in \S\ref{sec:example}.

In the second method, the data points with lower errors larger than
100\% have log space $\sigmal$ that is infinite, and therefore act as
upper limits and have no constraints on the models below the observed
fluxes.  Therefore we provide a third option so that certain
constraints can still be applied to the models even in such a
situation. If at some wavelength $\sigmal(\fnuobs)\geq 100\%~\fnuobs$,
we set $\sigmal(\log\fnufit)=\sigma_0\equiv2$. However, unlike the
normal data points, the contribution of this data point to the total
$\chisq$ is set to be $\mathrm{arcsinh}(x^2)=\ln (x^2+\sqrt{x^4+1})$,
where $x=(\log\fnumodav-\log\fnufit)/\sigma_0$, instead of $x^2$.  The
reason to choose such a function is that $\mathrm{arcsinh}(x^2)\simeq
x^2$ when $x$ is small and $\simeq \ln(2x^2)$ when $x$ is large, so
that unlike a data point with $\sigmal=\infty$ that has no constraint
on the model SED, it still tends to select the models with fluxes
closer to $\fnufit$, but the constraint is not as strong as a normal
data point with $\sigmal=\sigma_0$, since the contribution to the
$\chisq$ increases with $x$ in a logarithmic form. This method
  thus tends to select models that are themselves upper limits of the
  range of possibilities.

To sum up, when the errors of the observed fluxes are small and
symmetric, or the measurements and errors are statistically well
defined, the first option may be used.  Otherwise the second option
should be used. The third option provides a special treatment for the
data points with $>$100\% lower errors. In the fitting program,
different options can be assigned to different data points.

Here we emphasize that the $\chisq$ value we define here is not
  statistically meaningful, especially given the assumption of
  normally distributed errors, since in the more general case it is
  then not linked to a well-defined probability. Often we expect
  errors to be dominated by uncertain systematic effects, such as
  clump background subtraction. Thus the method presented here only
  provides a way to compare different models in the model grid to
  select the ones that are close to the given observed data.  That is
  to say, our focus is comparing the $\chisq$ values of models in one
  SED fitting, not on the absolute $\chisq$ values, nor on comparing
  the $\chisq$ values of fittings to different observations.

\section{G35.20-0.74 as an Example of SED Fitting}
\label{sec:example}

\subsection{Model Parameters and Degeneracies}
\label{sec:example_fit}

\begin{table*} 
\begin{center}
\caption{Parameters of the Five Best-fitting Models \label{tab:bestmodel}}
\ \\
\begin{tabular}{c|cc|ccc|ccc|ccccc}
\hline
\multirow{2}{*}{Model} & \multirow{2}{*}{$\chisq$} & \multirow{2}{*}{$\chisq_\mathrm{nonlimit}$} & $\mc^\mathrm{(a)}$ & $\scl^\mathrm{(a)}$ & $\ms^\mathrm{(a)}$  & $\inc^\mathrm{(b)}$ & $d^\mathrm{(b,c)}$ & $\av^\mathrm{(b)}$ & $R_c^\mathrm{(d)}$ & $\thetaw^\mathrm{(d)}$ & $\menv^\mathrm{(d)}$ & $\dot{m}_*^\mathrm{(d)}$ & $\ltot^\mathrm{(d)}$ \\
& & & ($M_\odot$) & ($\gcm$) & ($M_\odot$) & ($^\circ$) & (kpc) & (mag) & (AU) & ($^\circ$) & ($M_\odot$) & ($M_\odot~\mathrm{yr}^{-1}$) & ($L_\odot$) \\
\hline
1 & 2.64 & 3.12 & 480 & 0.1 & 16 & 48 & 2.2 & 42 & $1.1 \times 10^5$ & 15 & $4.4 \times 10^2$ & $1.2 \times 10^{-4}$ & $3.9 \times 10^4$ \\
2 & 2.70 & 3.90 & 100 & 3.16 & 12 & 34 & 2.2 & 26 & $8.5 \times 10^3$ & 20 & $7.7 \times 10^1$ & $9.4 \times 10^{-4}$ & $5.2 \times 10^4$ \\
3 & 2.84 & 3.35 & 200 & 0.316 & 12 & 22 & 2.2 & 45 & $3.8 \times 10^4$ & 17 & $1.7 \times 10^2$ & $1.9 \times 10^{-4}$ & $4.0 \times 10^4$ \\
4 & 2.90 & 3.43 & 320 & 0.1 & 24 & 68 & 2.2 & 82 & $8.6 \times 10^4$ & 27 & $2.6 \times 10^2$ & $1.2 \times 10^{-4}$ & $8.4 \times 10^4$ \\
5 & 3.12 & 3.69 & 400 & 0.1 & 16 & 58 & 2.2 & 38 & $9.6 \times 10^4$ & 17 & $3.6 \times 10^2$ & $1.1 \times 10^{-4}$ & $3.8 \times 10^4$ \\
\hline
\end{tabular}
\end{center}
Notes:\\
$^\mathrm{a}$ The primary parameters.\\
$^\mathrm{b}$ The additional independent parameters.\\
$^\mathrm{c}$ Fixed in this example.\\
$^\mathrm{d}$ Selected derived secondary parameters.\\
\end{table*}

We use the SED of the massive protostar G35.20-0.74 as an example for
demonstrating the SED fitting program with our model grid. The SED is
constructed based on the {\it SOFIA}-FORCAST observations from 20 to
40 $\mu$m, along with fluxes measured at other wavelengths from 3.6
$\mu$m to 500 $\mu$m using archived data of other instruments,
including {\it Spitzer}-IRAC, {\it Herschel}-PACS/SPIRE, and other
ground-based infrared instruments. The {\it SOFIA} observation of this
source is part of the {\it SOFIA} Massive Star Formation (SOMA) survey
(\citealt[]{Debuizer17}).  The SOFIA continuum images and SEDs of this
source were first presented and analyzed by \citet[]{Zhang13}, with
the then-under-development model grid and a limited, ad hoc
exploration of parameter space.  Later, the method used to derive the
SED from the continuum imaging was improved by \citet[]{Debuizer17}.
In that paper, the current model grid was used to fit the SED of
G35.20-0.74 and seven other massive protostars, but without detailed
discussion about the fitting process and the full results.  Here we
focus on using this SED as an example to demonstrate the SED fitting
program.  
The values of the fluxes and their errors are listed in \citet{Debuizer17}.
For the details about the observations and derivation of the
SED, please also refer to \citet[]{Debuizer17}.  Also, following
\citet[]{Debuizer17}, in the example here, we set the short wavelength
fluxes at $\lesssim 8~\mu$m to be upper limits, as discussed in
\S\ref{sec:caveat}.

\begin{figure}
\begin{center}
\includegraphics[width=0.5\columnwidth]{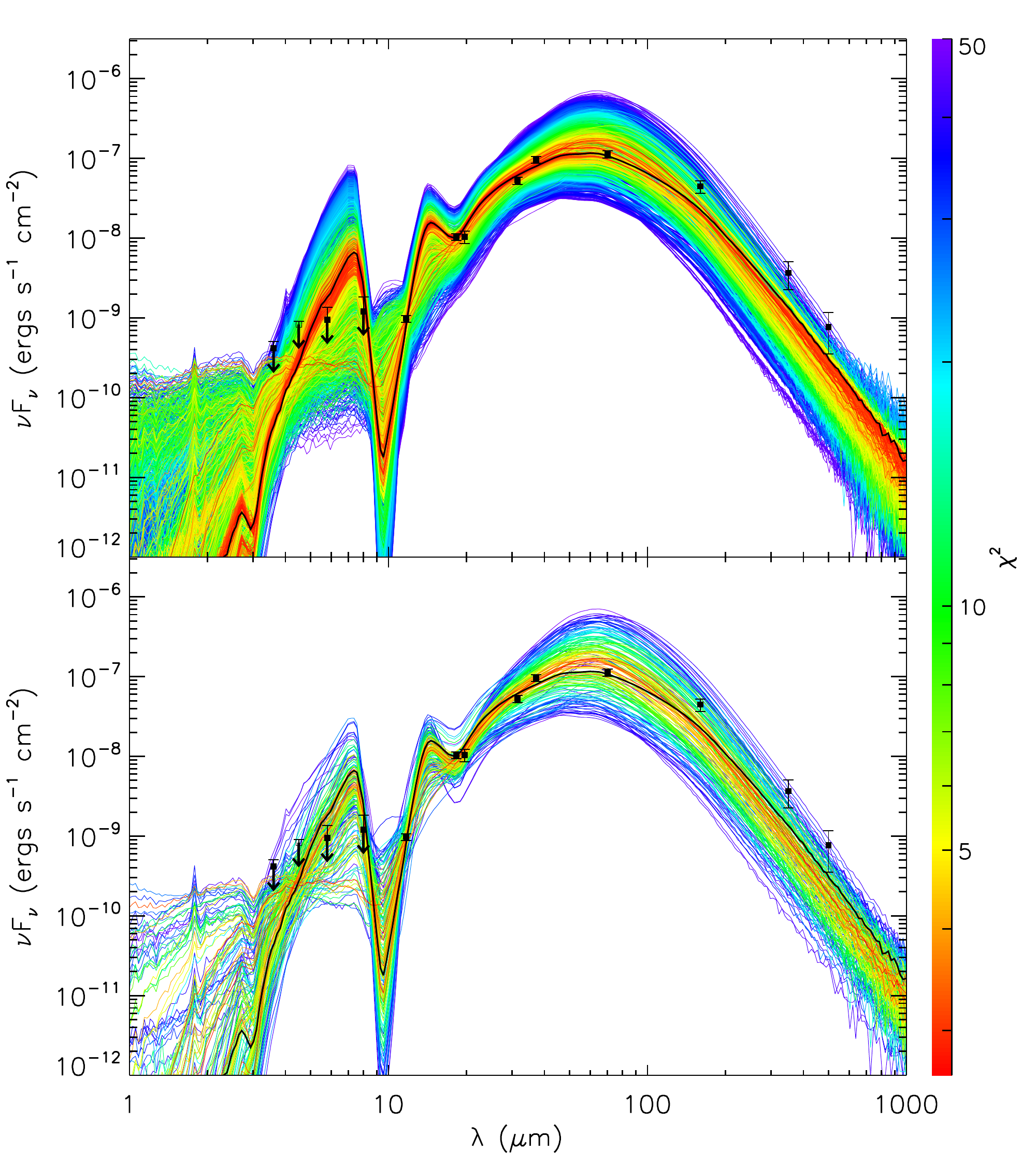}\\
\caption{SEDs of models with different $\chisq$ ($<50$) {\bf for the source G35.20-0.74}. 
The upper panel shows 
the model SEDs with different ($\mc, \scl, \ms, \inc$), and the lower panel shows
the model SEDs with different ($\mc, \scl, \ms$).  The thick black
line is for the best model. The black symbols are the observed fluxes
and errors. The upper limits are marked with arrows.}
\label{fig:sedrange}
\end{center}
\end{figure}

As described in the previous section, the fitting program produces two
groups of best models.  The first group contains models which are
different in the five dimensional parameter space comprised of $\mc$,
$\scl$, $\ms$, $\inc$ and $d$ (four dimensional parameter space if the
exact value of $d$ is provided).  The second group further selects the
best models that are different in the three dimensional space
comprised of the three primary parameters $\mc$, $\scl$, $\ms$.
Therefore the best models in the first group may share the same
physical model but differ only because of different distances or
viewing inclinations, while in the second group the models are
actually different in their initial conditions or evolutionary stages,
and have different physical structure and properties.  Figure
\ref{fig:sedrange} shows all the model SEDs 
with $\chisq<50$ in the two groups of results. Here the
distance to the source is set to be a fixed value of 2.2 kpc 
(\citealt[]{Zhang09}; \citealt[]{Wu14}).  
In this case, the best fit model has $\chisqmin=2.64$.
In the first group of results (upper panel), there are 171 model SEDs with
$\chisq-\chisqmin<3$, 670 model SEDs with $\chisq-\chisqmin<10$, and
2441 model SEDs with $\chisq<50$, among the total 8640 model SEDs. 
In the second group of results (lower panel), 
there are 27 models with $\chisq-\chisqmin<3$, 80 models with
$\chisq-\chisqmin<10$, and 178 models with $\chisq<50$, among the
total 432 model SEDs.  For the relatively well-fitted models with
$\chisq-\chisqmin<3$, on average, a change of 0.3 in $\cos\inc$ will
not affect the rank of the models determined by the primary parameters
of $\mc$, $\scl$ and $\ms$.  We can consider $\pm0.15$ in the
$\cos\inc$ space to be the fitting uncertainty of the inclination in
this case.  But as we will show below, this is actually dependent on
the exact model.

As discussed in the previous section, in our fitting, the upper
  limits work as a normal data points and contribute to $\chisq$ if
  the model SED is higher than the upper limits. Therefore, it is
  possible that the best models are above one or more upper limits,
  which is the case shown here.  However, users can assign smaller
  errors to the upper limits to make them stronger constraints.  In
  fact, very small errors for the upper limits practically exclude
  models with fluxes above the upper limits.  It is also worth noting
  that, in this example, the group of best models also appear to
  under-predict the fluxes at longer wavelengths $\gtrsim160~\mu$m.
  Among the 7 sources whose SEDs were fitted with our model grid by
  \citet[]{Debuizer17} (8 sources were fitted in this paper, but one
  source is without fluxes at wavelengths $>100~\mu$m; see also
  Section \ref{sec:example_SOMA}), two sources (including G35.2-0.74)
  show such deviation at the long wavelengths.  Since the slope at
  wavelengths $>200~\mu$m of the model SEDs are similar to the
  observed slopes, the adopted dust models are not likely to be the
  reason for such deviation.  It is more likely caused by the
  inclusion of additional clump material at these long wavelengths
  when deriving the SEDs from observations.

\begin{figure}
\begin{center}
\includegraphics[width=\columnwidth]{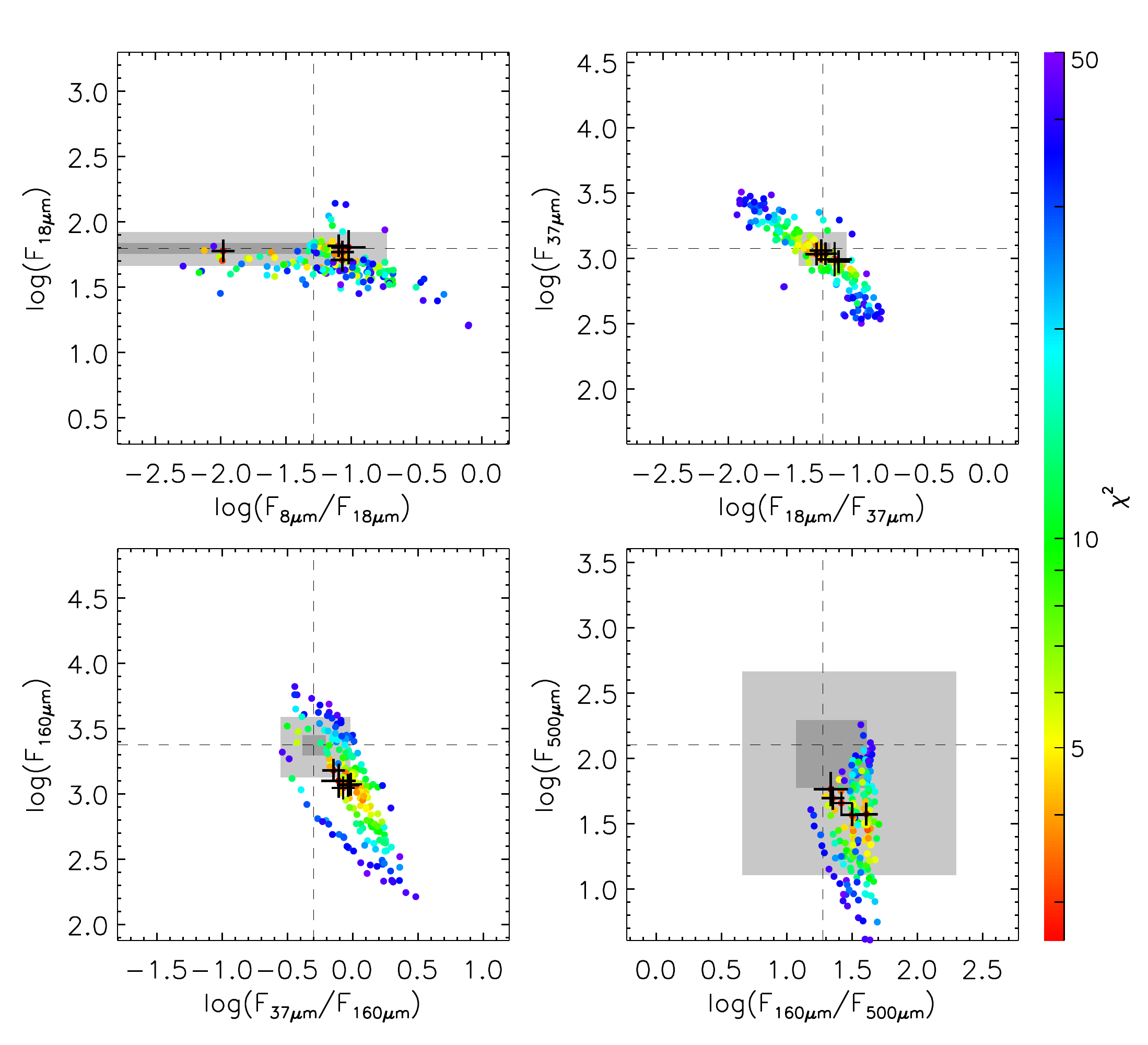}\\
\caption{The distribution of $\chisq$ in the color-flux space around selected
wavelengths, showing how the models deviate from the observations at
these wavelengths.  The center of each panel is the location of the
observational data, and the dark and light grey areas show the ranges
of 1$\sigma$ and 3$\sigma$.  Only the models with the $\chisq<50$ are
shown.  Each model is different in the ($\mc, \scl, \ms$)
space.  The crosses mark the locations of the best five models, and
the large cross is the best model.}
\label{fig:plotchisq_color}
\end{center}
\end{figure}

\begin{figure}
\begin{center}
\includegraphics[width=0.5\columnwidth]{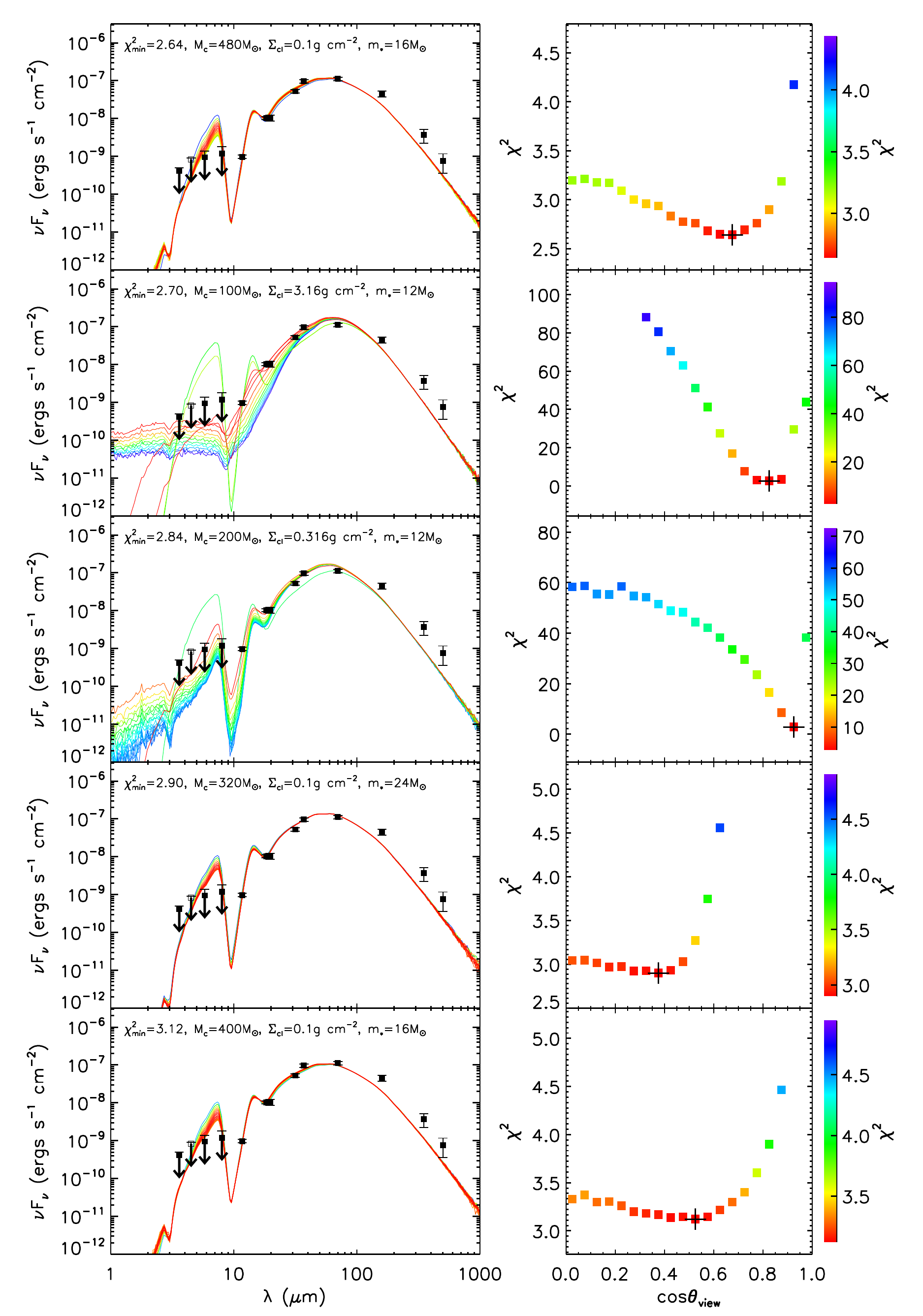}\\
\caption{{\bf Left:} The panels show the SEDs with same $\mc, \scl, \ms$
as the five best models (from top to bottom in increasing order of $\chisq$; see Table \ref{tab:bestmodel}) 
but different inclinations. At each inclination, $\av$ is adjusted to minimize the $\chisq$. 
{\bf Right:} The panels show how the inclination would change $\chisq$ for the five best models.
The best models are marked with crosses.}
\label{fig:sedrange1}
\end{center}
\end{figure}

\begin{figure*}
\begin{center}
\includegraphics[width=0.93\textwidth]{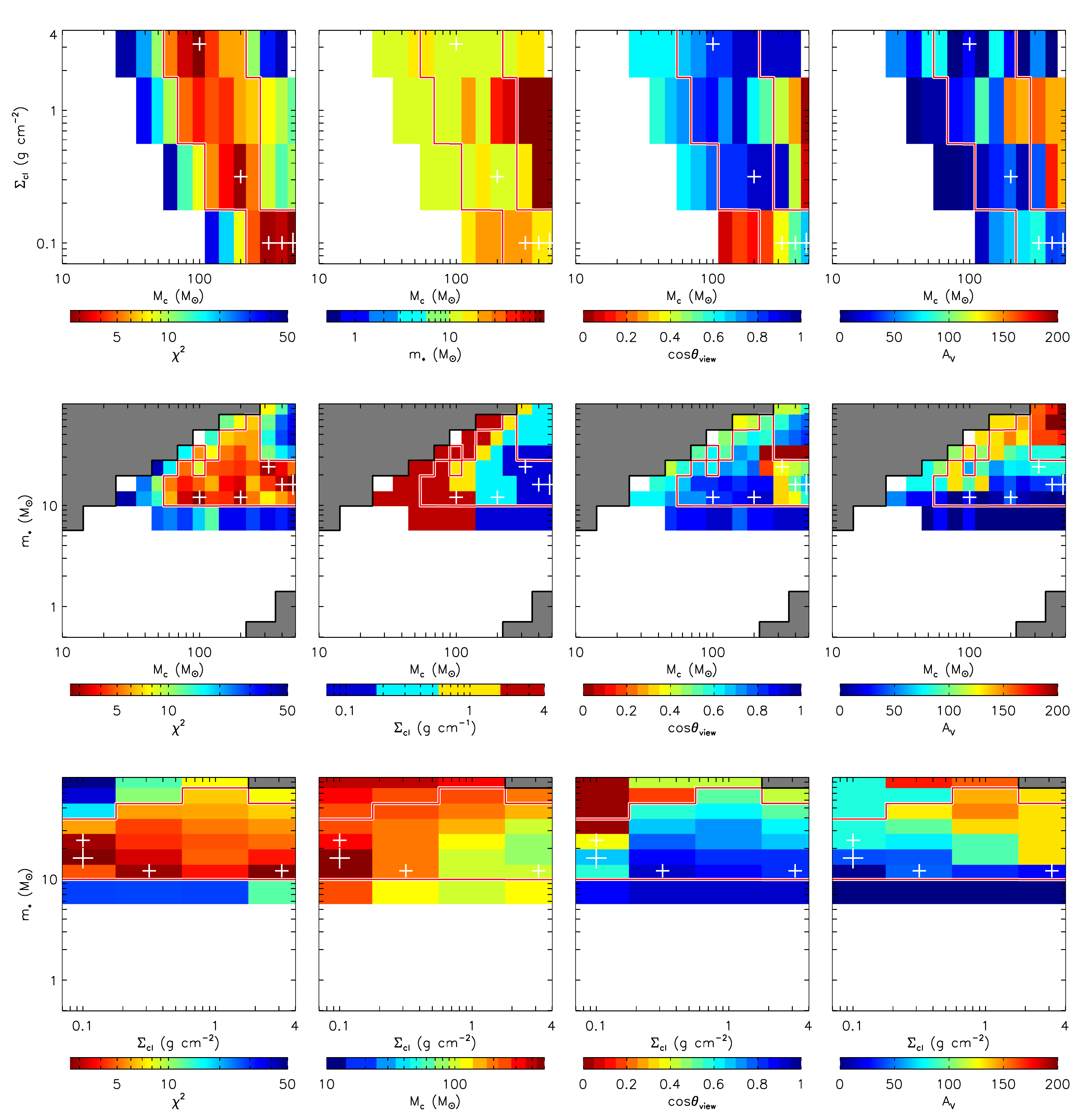}\\
\caption{The distribution of $\chisq$ in the parameter space of $(\mc, \scl,
\ms)$.  {\bf Top row}: The first panel shows the best $\chisq$ for
each set of $(\mc, \scl)$, by searching through different $\ms$,
$\inc$ and $A_V$. The $\ms$, $\inc$ and $A_V$ that are used to achieve
these best $\chisq$ are shown in the second to fourth panels of this
row.  {\bf Middle row}: Similar to the top row, but shows the best
$\chisq$ in $\mc-\ms$ space, and the $\scl$, $\inc$, $A_V$ to achieve
these best $\chisq$.  {\bf Bottom row}: Similar to the top and middle
rows, but shows the best $\chisq$ in $\scl-\ms$ space. The white
crosses mark the locations of the five best models, and the large
cross is the best model.  The grey regions are not covered by the
model grid, and the white regions are where the $\chisq>50$. The
red contours are at the level of $\chisq=\chisq_\mathrm{min}+5$.}
\label{fig:plotchisq_primary}
\end{center}
\end{figure*}

\begin{figure*}
\begin{center}
\includegraphics[width=\textwidth]{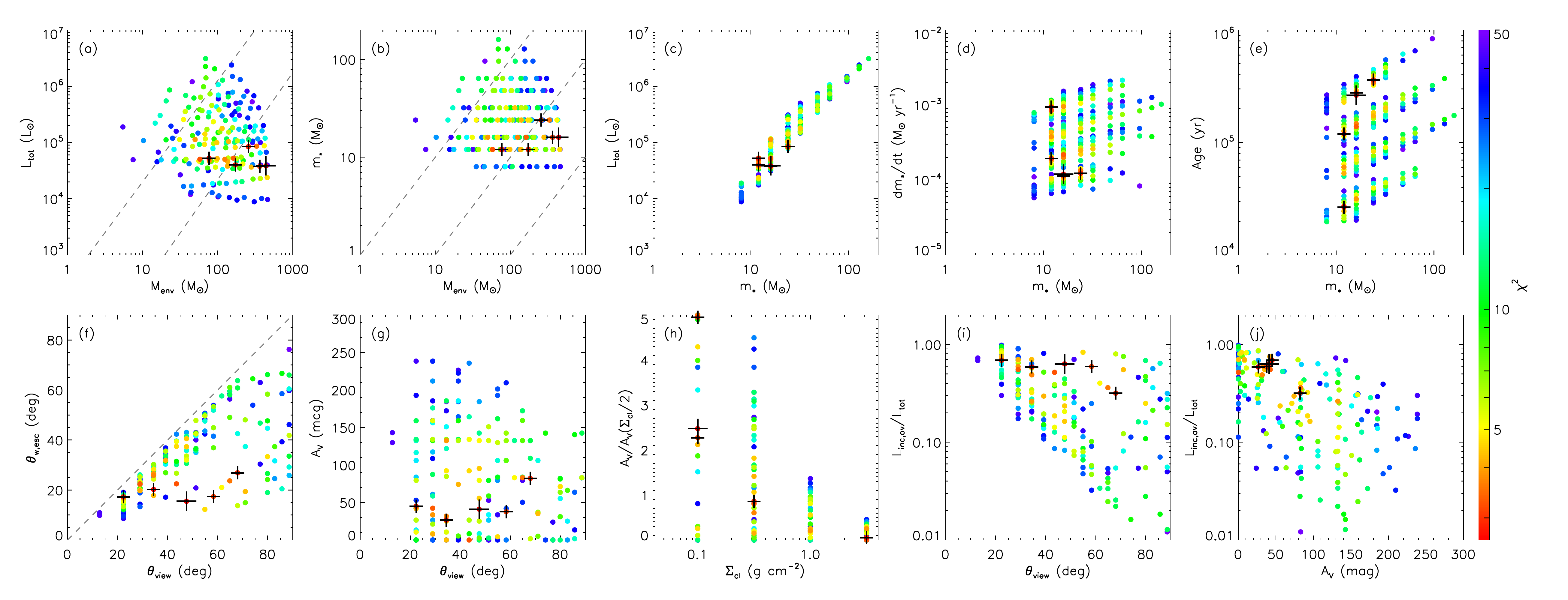}\\
\caption{The distribution of $\chisq$ with various secondary parameters (see the text for details). 
Only the models with the $\chisq<50$ are shown.  Each model is different in
the ($\mc, \scl, \ms$) space.  The crosses mark the location of the
best five models, and the large cross is the best model.
The dashed lines in Panel (a) are the fitted $\ltot-\menv$ relation for models around $\ms/\menv\approx 0.1$
and $\ms/\menv\approx 1$ discussed in \S\ref{sec:parameter}. 
The dashed lines in Panel (b) are $\ms/\menv=0.01$, 0.1, and 1.
The dashed line in Panel (f) is where $\inc=\thetaw$.}
\label{fig:plotchisq_secondary}
\end{center}
\end{figure*}

\begin{figure}
\begin{center}
\includegraphics[width=0.5\columnwidth]{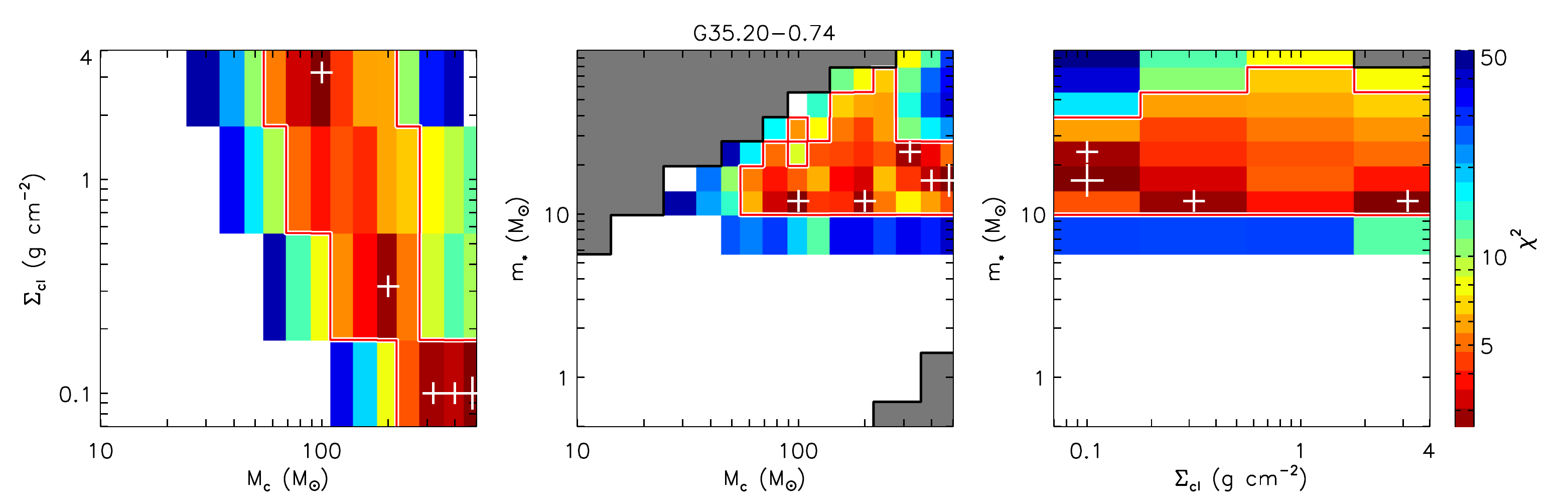}\\
\includegraphics[width=0.5\columnwidth]{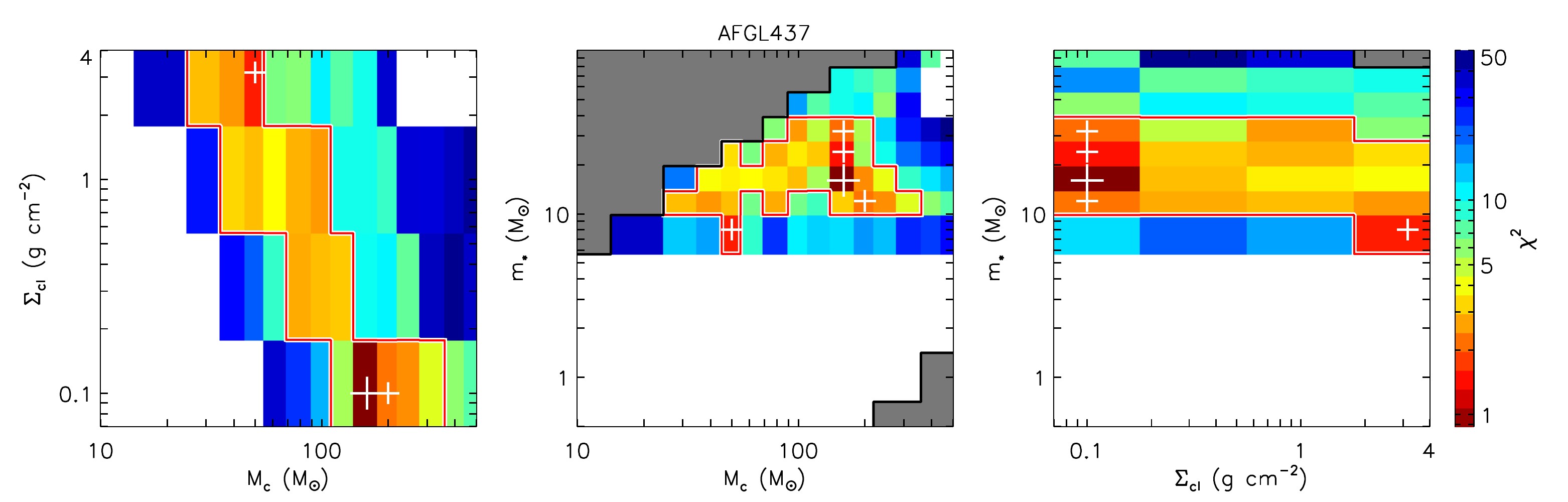}\\
\includegraphics[width=0.5\columnwidth]{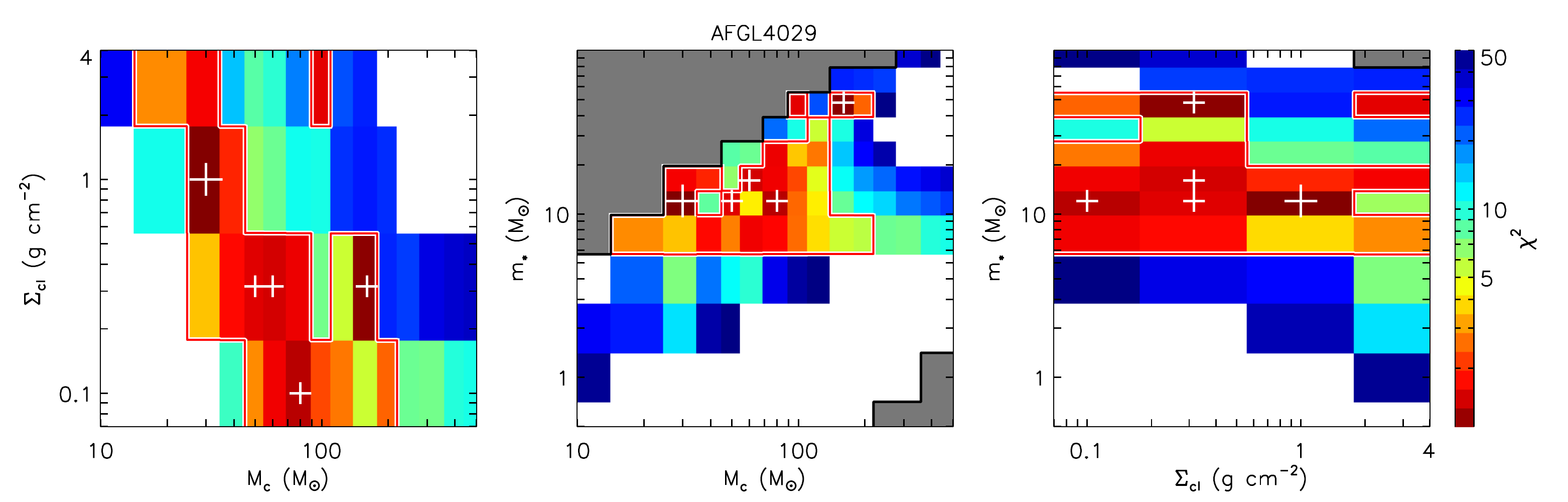}\\
\includegraphics[width=0.5\columnwidth]{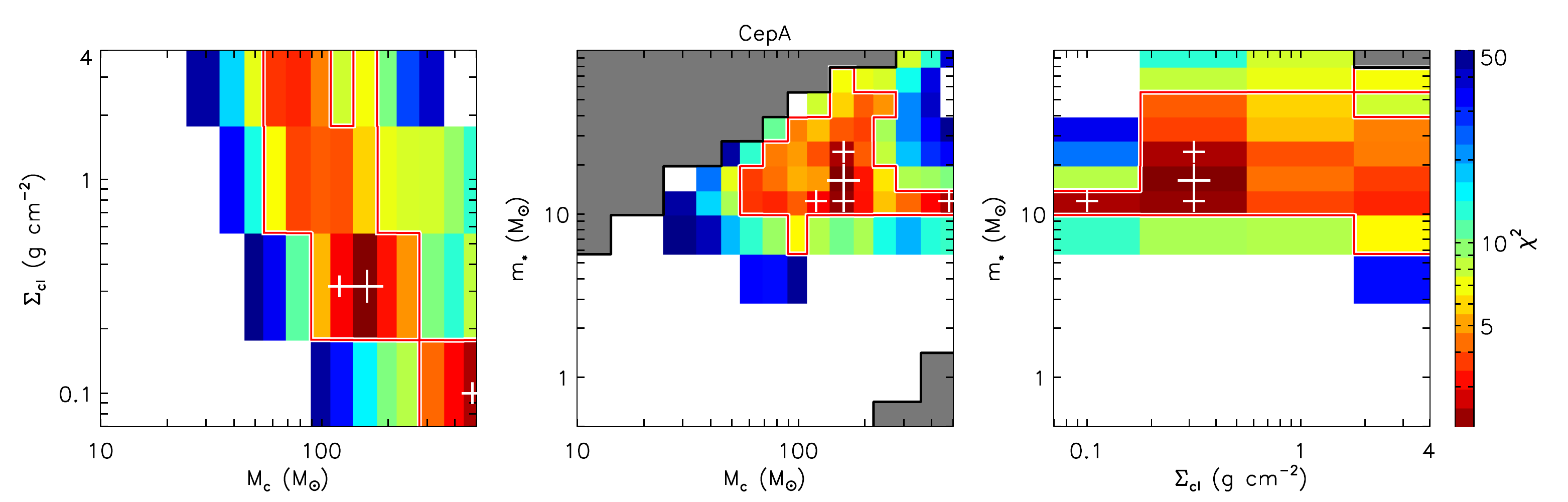}\\
\includegraphics[width=0.5\columnwidth]{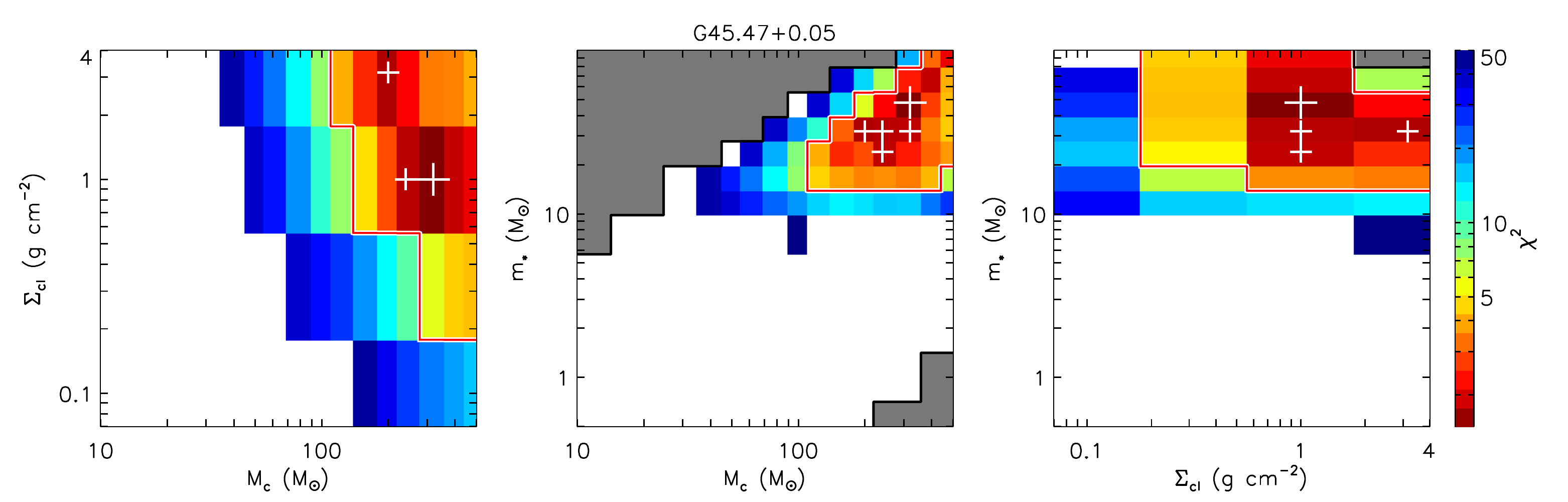}\\
\includegraphics[width=0.5\columnwidth]{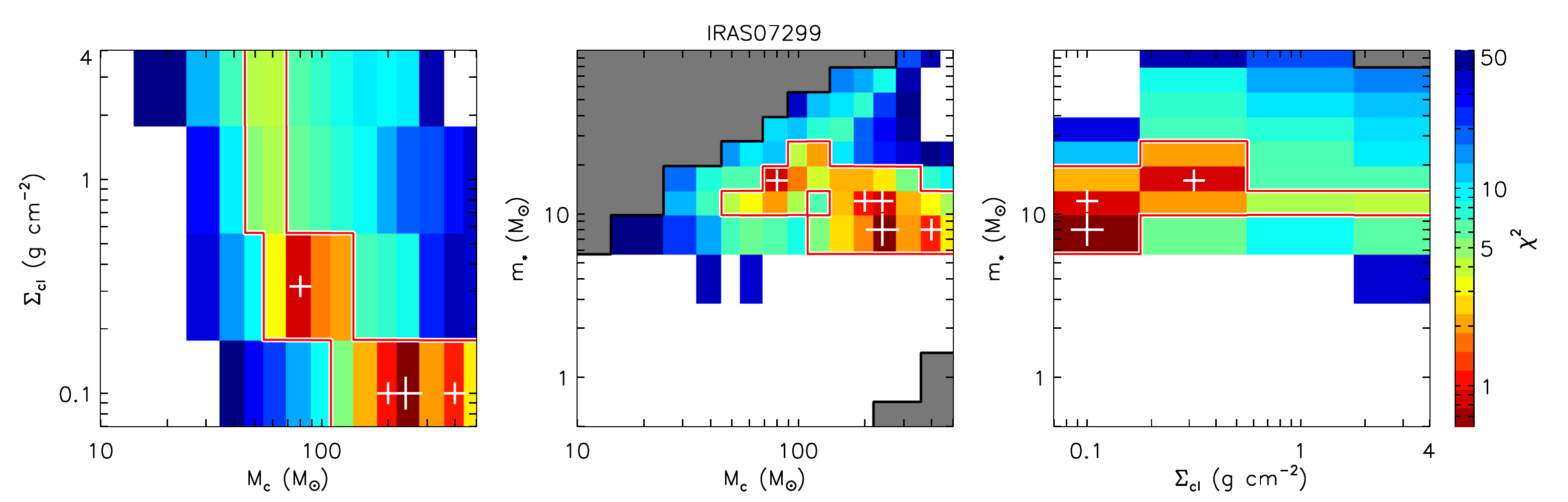}\\
\includegraphics[width=0.5\columnwidth]{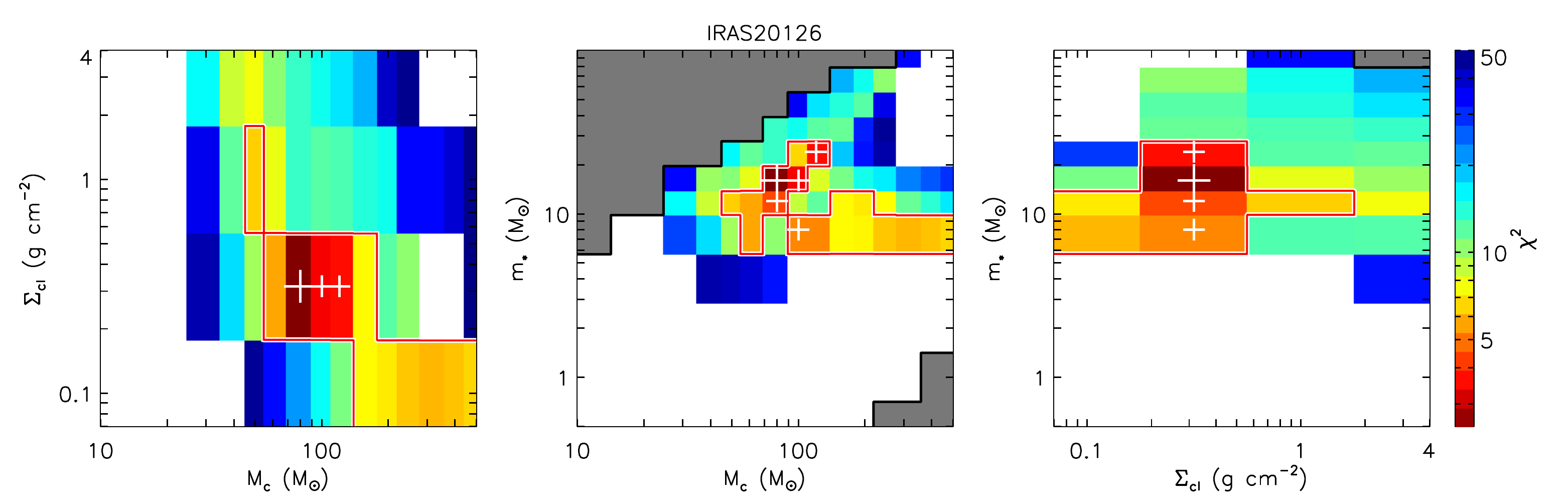}\\
\includegraphics[width=0.5\columnwidth]{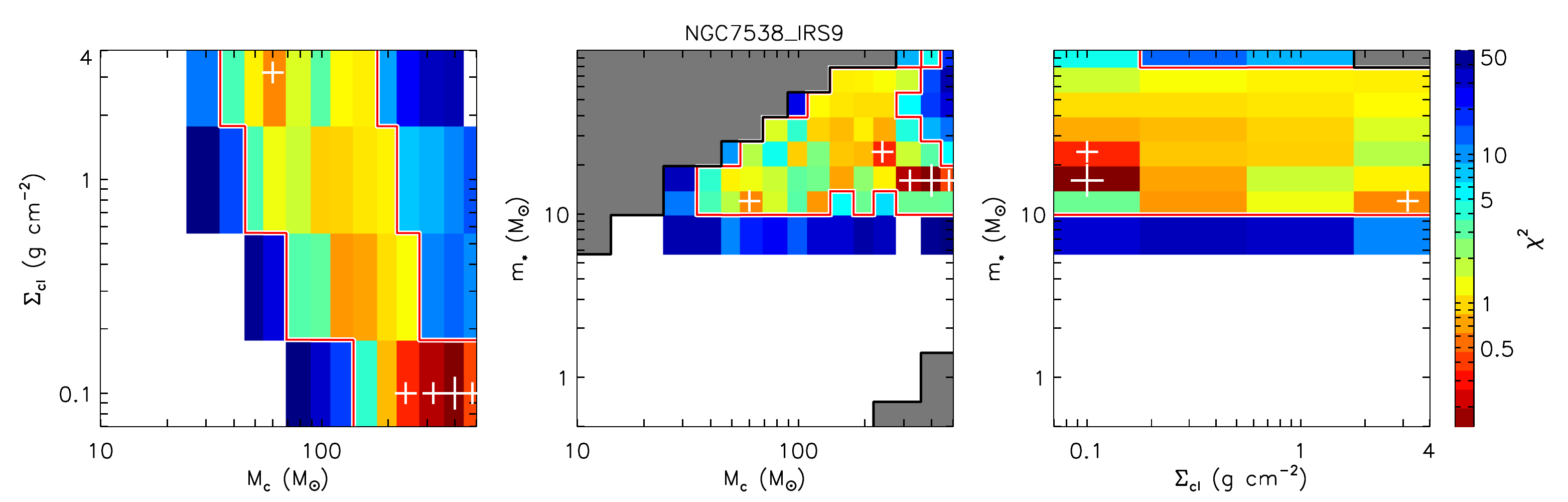}\\
\caption{Short-format SED model fitting outputs for the eight sources of the
SOMA survey (\citealt[]{Debuizer17}). See this paper for detailed
discussion of the properties of these protostars. The panels are similar to
those in the first column of Figure \ref{fig:plotchisq_primary}.}
\label{fig:plotchisq_primary_8source}
\end{center}
\end{figure}

Figure \ref{fig:plotchisq_color} shows how the models deviate from the
observed fluxes and colors (i.e., SED slopes) at different
wavelengths. Only the models in the second group of results are
shown. The best five models are also marked.  Most of the models with
$\chisq-\chisqmin<3$ ($\chisqmin=2.64$), which corresponds to an average deviation of
$<2.4\sigma$ at each data point, are indeed located within the range
of $\lesssim 2\sigma$ at the wavelengths shown here, suggesting the
data points of different wavelengths are evenly contributing to the
$\chisq$, and the fitting is constrained over the whole range of
wavelengths. The fitting is especially well constrained in the
wavelength range of $20-40~\mu$m in terms of the absolute fluxes and
also the SED slope.  The best models are under-predicting the
$160~\mu$m flux at levels of about $3\sigma$, and also
under-predicting the $500~\mu$m flux at levels of about $2\sigma$.  In
addition, most of the models are predicting a slightly steeper slope
between 160 and 500 $\mu$m at levels within $1\sigma$.  At short
wavelengths, although the data points at $8~\mu$m are set to be upper
limits, most of the best fitted models are still close to the data
points.

Table \ref{tab:bestmodel} lists the parameters of the five best models
in the second group of results which are different in the primary
parameter space of ($\mc$, $\scl$, $\ms$).  The additional independent
parameters ($\inc$, $d$ and $\av$) to achieve these minimum $\chisq$
are also listed.  We also list several important parameters that are
derived from the primary parameters based on the physical model,
including: the core radius $R_c$, which in the model only depends on
the initial conditions $\mc$ and $\scl$ and is constant over time; the
opening angle of the outflow cavity $\thetaw$; the current envelope
mass $\menv$; the accretion rate from disk to protostar $\dot{m}_*$;
and the total luminosity $\ltot$. Note that $\ltot$ is different from
the value directly integrated from the SED due to the effect of
inclination (i.e., the flashlight effect) and foreground extinction.
Besides the $\chisq$ used to rank the models,
$\chisq_\mathrm{nonlimit}$ is also listed, showing the average
deviation between the model SEDs and the observed data points.
  Note these results are slightly different (slight change of ordering
  of the best models) from those shown by \citet[]{Debuizer17}.  This
  is because here we have improved the quality of the
  model SEDs by using larger number of photon packets in the
  Monte-Carlo RT simulations and reducing the Monte-Carlo noise levels
  of the model SEDs.  The improvement of the model SED qualities
  slightly affects the fitting especially at short-wavelengths for
  more embedded or more edge-on sources.  Among the best five models,
the initial core mass $\mc$ is constrained to be $\gtrsim
100\:M_\odot$ (most likely $\gtrsim 200\:M_\odot$) and the
protostellar mass is constrained to be $10-20\:M_\odot$.  The half
opening angle of the outflow cavity is constrained to be
$15-30^\circ$, the accretion rate is constrained to be between
$10^{-4}$ to $10^{-3}\:M_\odot\:\mathrm{yr}^{-1}$ (most likely
$(1-2)\times 10^{-4}\:M_\odot\:\mathrm{yr}^{-1}$), and the total
luminosity is about $(4-8)\times 10^4\:L_\odot$.  On the other hand,
the clump environment mass surface density $\scl$ and the inclination
angle are not well constrained.

For these five best models, Figure \ref{fig:sedrange1} shows how the
inclination would change the fitting when other parameters are kept
the same. Note that a different foreground extinction, $\av$, is
adopted to minimize the $\chisq$ for each inclination angle.  For
three of the five models (Models 1, 4 and 5 shown in 
the first, fourth and fifth rows; see also Table \ref{tab:bestmodel}),
the fitting results are
not so sensitive to inclination, except close to $\cos\inc=1$, i.e., a
face-on view.  In this case the inclination is not well constrained.
On the other hand, in the other two models (Models 2 and 3), the
fitting results are highly sensitive to the inclination angle.  The
reason for this difference is that in the former case, the unextincted
SEDs of different inclinations are all above the observed data points
at wavelengths $\lesssim 70~\mu$m, and by adjusting foreground
extinction $\av$, the model SEDs of different inclinations except
close to a face-on view can all have relatively good fits to the
observations.  In the latter case, the unextincted model SEDs of
higher inclinations are below the observed fluxes and the fitting
cannot be improved by adjusting $\av$, therefore relatively good
fitting can be achieved only at a narrow range of inclinations.  This
suggests a certain degeneracy between the inclination $\inc$ and
foreground extinction $\av$.

Figure \ref{fig:plotchisq_primary} shows the distribution of the
models in the primary parameter space comprised of $\mc$, $\scl$, and
$\ms$.  The first row shows the distribution of $\chisq$ of the best
models in the $\mc-\scl$ space in the first panel, and the
distributions of $\ms$, $\inc$, $\av$ to achieve these best models in
the second to fourth panels.  Similarly, the second row shows the
distribution of $\chisq$ of the best models in the $\mc-\ms$ space and
the corresponding distributions of $\scl$, $\inc$ and $\av$, and the
third row shows the equivalent distributions in the $\scl-\ms$ space.
In the $\mc-\scl$ space, the models with $\chisq-\chisqmin<5$ (inside
the red contour) occupy a region with high $\mc$ but spanning the full
range of $\scl$ (first row, first panel).  For these models, the
initial core mass $\mc$ appears to be higher in a lower surface
density environment.  The range of $\mc$ of the best models
($\chisq-\chisqmin<5$) gradually increases from $60-200\:M_\odot$ at
$\scl=3.2\:\gcm$ to $240-480\:M_\odot$ at $\scl=0.1\:\gcm$.  The
models are more constrained in $\ms$ (e.g., second row, first panel).
All of the best models ($\chisq-\chisqmin<5$) have $\ms$ higher than
about 10~$M_\odot$ and mostly have $\ms\leq 24\:M_\odot$, but extend
to higher $\ms$ for the models with $\mc$ in the range of
$160-240\:M_\odot$.  The fitted inclinations are mostly around
$\cos\inc\approx 0.8$, i.e., $\inc\approx 37^\circ$ between the line
of sight and the outflow axis, for the best models with
$\scl=0.3-3\:\gcm$ and $\mc=60-200\:M_\odot$ (e.g., first row, third
panel).  For the best models with low $\scl$ ($0.1\:\gcm$) and high
$\mc$ ($>200\:M_\odot$), $\inc$ is in the range of about
$50^\circ-80^\circ$.  For most of these best models, a foreground
extinction of $\av<100$~mag is needed to achieve the minimum $\chisq$
(e.g., first row, fourth panel).  To summarize, in the example shown
here, while $\mc$ and $\ms$ are relatively well constrained, $\scl$ is
not. However if the inclination can be independently determined from
other observations such as those of outflow kinematics, then this
degeneracy can be further broken.

Figure \ref{fig:plotchisq_secondary} shows the distribution of all the
second group models (different in the $\mc-\scl-\ms$ space) with
$\chisq<50$ with various secondary parameters. In the model grid,
while all the SEDs are determined by the three primary parameters
$\mc$, $\scl$, and $\ms$, along with the additional independent
parameters $\inc$, $d$ and $\av$, the secondary parameters, which are
derived from the primary parameters and describing the properties of
the protostar, envelope, outflow cavity and disk, are of more interest
in understanding the physical conditions of the observed massive
protostars. Panels (a) and (b) show the distribution of the models in
the $\menv-\ltot$ and $\menv-\ms$ diagrams.  As discussed in previous
sections, the location in the $\menv-\ltot$ diagram and the parameter
$\ms/\menv$ are often used as indicators of the evolutionary
stage. Most of the models with $\chisq<50$ are located between
$\ms/\menv=0.1$ and 1, which corresponds to the turning points of the
evolutionary tracks in the $\menv-\ltot$ diagram (see Figure
\ref{fig:masslum}), and the best models that have $\chisq-\chisqmin<3$
(red and orange colors) are located around $\ms/\menv=0.1$. This
suggests that this source is still highly embedded and in the main
accretion phase.

Panels (c), (d) and (e) show the relation of the total luminosity
$\ltot$, accretion rate $\dot{m}_*$, and age since the start of star
formation, with the protostellar mass $\ms$ in the fitted
models. These relations are determined by the evolutionary models (see
Figure \ref{fig:history}).  The total luminosity is highly dependent
on the protostellar mass and not so much on the other primary
parameters.  Since $\ms$ is relatively well constrained as discussed
above, the total luminosity is also constrained well.  The best models
with $\chisq-\chisqmin<3$ have a luminosity from several $\times 10^4$
to $\gtrsim 10^5~L_\odot$.  The accretion rates 
of the models with $\chisq-\chisqmin<3$ are constrained to be 
$10^{-4}-10^{-3}~M_\odot~\mathrm{yr}^{-1}$, and the protostellar age 
from several $\times 10^4$ to several $\times 10^5$ yr.
They are affected by the environmental mass
surface density $\scl$, which is less well constrained in this case.

Panel (f) shows the relation between the inclination and the opening
angle of the outflow cavity of the fitted models.  All the models have
an inclination angle larger than the outflow cavity opening angle,
i.e., the line of sight toward the protostar goes through the
envelope. As discussed above, the inclination angle is not well
constrained in this case, but interestingly, for the models with
$\chisq-\chisqmin<3$, the opening angles are either close to the
inclination angle up to about 40$^\circ$ or around $20^\circ$, despite
the wide range of inclinations.  All the best five models have opening
angles around $20^\circ$.

Panel (g) shows the distribution of the fitted models in the space of
the two additional independent parameters $\inc$ and $\av$. The models
with $\chisq-\chisqmin<3$ have $\av\lesssim 150$~mag, with most of
them within 100~mag. Panel (h) further compares the fitted $\av$ with
the values that correspond to the mean mass surface densities of the
ambient clumps $\scl$. On average, the contribution of the ambient
clump to the foreground extinction to the core should correspond to
$\scl/2$, which we define as $A_{V,\scl/2}$. But the foreground
extinction in the real situation will differ from $A_{V,\scl/2}$ due
to clumpy and/or anisotropic structures in the ambient clump, and
additional foreground extinction which is not related to the host
star-forming clouds.  For the best fitted models which have
$\chisq-\chisqmin<3$, with $\scl=0.3 - 3~\gcm$, the fitted $\av$ is
within $2A_{V,\scl/2}$, but for models with $\scl=0.1$, the fitted
$\av$ can be as high as $5A_{V,\scl/2}$, which indicates that the
foreground extinction needed to fit the observed SED may not be only
that expected from the ambient clump.
Thus, if a constraint is imposed that the foreground extinction should
be no more than that expected given the value of $\scl$, then some low
$\scl$ models would be excluded in this case.

Panels (i) and (j) compare the fitted total luminosities, $\ltot$,
with $L_{\mathrm{inc},A_V}$, the bolometric luminosities directly integrated from
the observed SED (see also Figure \ref{fig:flashlight}).  The latter luminosity is also slightly dependent on
the model fitting because of the uncertainties of the SED at the
wavelength ranges not covered by the observation.  For the models with
$\chisq-\chisqmin<3$, $L_{\mathrm{inc},A_V}/\ltot$ is in the range of
$0.2-0.8$.  This is caused by a combined effect of the inclination
(flashlight effect) and the foreground extinction (see
\S\ref{sec:flashlight}). According to Figure \ref{fig:flashlight},
such a ratio between $L_{\mathrm{inc},A_V}$ and $\ltot$ is consistent
with an outflow opening angle of about $20^\circ-30^\circ$.

In the end, we note that the above discussion is based on a specific
example and may not be general (see \S\ref{sec:example_SOMA}).  However, during each fitting, in
addition to giving just a few best models, the program will generate
similar figures to help the users to better understand the
results. The above discussion serves as an example of what information
we can expect from these figures.

\subsection{Discussion of the Source}
\label{sec:example_source}

G35.20-0.74 is a massive protostar in a broader region of star
formation located at a distance of 2.2 kpc (\citealt[]{Zhang09};
\citealt[]{Wu14}). CO(1-0) and (2-1) observations have revealed a wide
outflow structure in the direction of northeast-southwest which
extends to $>1\arcmin$ from the central source (\citealt[]{Gibb03};
\citealt[]{Birks06}).  Perpendicular to this, CS(2-1) observations
have revealed a ridge-like structure (\citealt[]{Dent85}), which has
been further resolved into a $15\arcsec$-long filament with a string
of cores embedded by ALMA (\citealt[]{Sanchez13,Sanchez14}) and SMA
(\citealt[]{Qiu13}) in the sub-mm continuum.  \citet[]{Heaton88}
observed this source in centimeter radio continuum and were able to
resolve three compact sources arranged north-south, and concluded that
the central source was likely an UCH$_\mathrm{II}$ region, while the
north and south sources had spectral indices consistent with free-free
emission from a collimated, ionized, bipolar jet. Since then it has
been debated whether the NE-SW CO outflow is caused by the ionized jet
undergoing precession, or if they are composed of separate outflows driven by
different sources.  This elongated radio continuum emission was
further resolved into 17 individual knots lying along the N-S
direction by the VLA (\citealt[]{Beltran16}; \citealt[]{Gibb03}), with
the driving source identified as an UC/HCH$_\mathrm{II}$ region.
This radio source is coincident with one of the embedded cores
identified in sub-mm observations (Core B identified by
\citealt[]{Sanchez13,Sanchez14} and MM1b identified by
\citealt[]{Qiu13}).  The N-S outflow is also seen in NIR and MIR
observations (\citealt[]{Dent85b}; \citealt[]{Debuizer06}).  At these
wavelengths, the emission is elongated in the N-S direction but peaked
to the north of the identified radio source and continuum core. It has
been argued that the outflow/jet is blue-shifted to the north and the
emissions at NIR and MIR are dominated by the northern near-facing
outflow cavity.  However, at longer wavelengths of $30-40~\mu$m, the
SOFIA-FORCAST observations have revealed the southern, far-facing
outflow cavity (\citealt[]{Zhang13}).

In our previous fitting of the SED of this source
(\citealt[]{Zhang13}) with an earlier version of our radiative
transfer models (which had fixed outflow cavity-opening angles) and
using a limited, ad hoc exploration of model parameter space, this
source was estimated to be a protostar with $\ms\simeq 20-34~M_\odot$,
accreting at rates of $\dot{m}_*\simeq
10^{-4}~M_\odot~\mathrm{yr}^{-1}$. The total luminosity was estimated
to be $(7-22)\times 10^4\:L_\odot$, and the opening angle of the
outflow cavity was estimated to be $35^\circ-50^\circ$.  Compared with
these earlier results, our completed model grid and SED fitting
program presented in this paper have estimated a protostellar mass of
$\ms=10-20\:M_\odot$ and a total luminosity of $(4-8)\times
10^4\:L_\odot$.  \citet[]{Sanchez13} has identified a Keplerian disk
in Core B with rotation corresponding to a central mass of
$18\:M_\odot$, and they argued that the disk is around a binary based
on the total luminosity. Indeed, a binary system of UC/HCH$_\mathrm{II}$ regions 
is seen by \citet[]{Beltran16} at the position of Core B.
Our model is based on a single protostar,
which under-predicts the total protostellar mass if the luminosity is
from a binary, therefore our new estimation is quite consistent with
the mass estimation from gas kinematics.  Compared with the previous
estimation, our new estimation of the opening angle of the outflow
($\sim 20^\circ$) is also more consistent with the MIR observations
which suggests a narrow outflow cavity (e.g., \citealt[]{Debuizer06}).

Our SED fitting also estimates the current envelope mass to be about
$100-400~M_\odot$, and a ratio between the protostellar mass to
envelope mass of $\ms/\menv\approx 0.1$. The total mass of the
filament has been estimated to be about 160 $M_\odot$ and the mass of
the core that hosts the driving source of the N-S outflow/jet was
estimated to be about 18 $M_\odot$ (\citealt[]{Qiu13}).  However,
these masses are concentrated to the fragments with sizes of about
$1\arcsec$, and the MIR emission shows that there is a narrow outflow
cavity existing on a scale of about $10\arcsec$, suggesting envelope
material extends at least to such a scale.  This indicates that the
total mass of the gas envelope surrounding this N-S outflow should be
higher than 160 $M_\odot$, which is consistent with our estimation of
the current envelope mass.  However, unlike our model which has a
highly idealized spherical core with smooth density distribution, the
observations show that the envelope is actually highly fragmented and
may be feeding several protostars and close binary systems.
Therefore, if one of the protostars or close binary systems is
contributing most of the IR emission (e.g., Core B in this case), our
model grid is still able to generate relatively accurate estimates
about this main source and overall properties of the large
envelope. However, due to the fragmentation, the mass reservoir for
each protostar or close binary system is smaller than what the model
suggests and therefore the sources may be at a later evolutionary
stage than indicated in the models.

We note also that our previous model fitting of G35.20-0.74
\citep{Zhang13} used not only the SED, but also the MIR to FIR flux
intensity profiles along the outflow axis. In a follow-up paper, we
plan to extend our presented model grid to include multiwavelength
images, which can be used in such ways to further constrain the
protostellar properties.

\subsection{Standard Format Output for SOMA Survey Protostars}
\label{sec:example_SOMA}

Finally, to illustrate simplified outputs from the SED fitting tool,
in Figure \ref{fig:plotchisq_primary_8source} we show a standard
format output of 3 panels, i.e., $\scl$ vs. $\mc$, $\ms$ vs. $\mc$,
and $\mc$ vs. $\scl$, which were already presented in Figure
\ref{fig:plotchisq_primary}. These results are shown for G35.20-0.74,
and also for seven other massive protostars from the first SOFIA
Massive (SOMA) Star Formation survey data release
(\citealt[]{Debuizer17}). We do not discuss these other sources
individually in detail here, but these results complement the simpler
SED fitting results (i.e., the lists of the top 5 SED models) for
these sources presented by \citealt[]{Debuizer17} and the discussion of
these sources in this paper.

Concerning general trends, from Figure
\ref{fig:plotchisq_primary_8source} we see that among the three
primary parameters, the protostellar mass $\ms$ is best constrained.
Most of the sources have $\ms$ around $10-20~M_\odot$. But in
G45.47+0.05 $\ms$ is clearly higher around $30-40~M_\odot$ and
in IRAS 07299, $\ms$ is slightly lower (around $8-16~M_\odot$).  In the $\mc-\scl$
space, for all the sources, the best models ($\chisq-\chisqmin<5$,
within the red contours) occupy a region with lower $\mc$ at higher
$\scl$ and higher $\mc$ at lower $\scl$, similar to what we found in
G35.20-0.74 in the previous sections.  But there is a clear difference
in the ranges of $\mc$ of the best models from source to source, with
G45.47-0+0.05 having highest $\mc$ and AFGL4029 having relatively low
$\mc$.  As discussed above $\scl$ is least constrained, with most of
the sources having best models spanning over the full range of $\scl$
in the model grid. The constraint of $\scl$ is slightly better in IRAS
20126 and IRAS 07299 in which the best models concentrate in
$\scl=0.1-0.3~\gcm$, and G45.47+0.05, where the best models
concentrate in $\scl\gtrsim 1~\gcm$.

\section{Discussion and Conclusions}
\label{Summary}

We have presented a model grid for fitting the SEDs of massive
protostars. The model grid is based on the Turbulent Core model of
massive star formation (\citealt[]{MT02, MT03}). The initial conditions of the
model grid are pressurized, dense, massive cores embedded in high mass
surface density ``clump'' environments. These initial conditions are
parameterized by the initial mass of the core, $\mc$, and the mean
mass surface density of the clump, $\scl$.  Using analytical and
semi-analytical solutions, we self-consistently calculate the
properties and evolutions of the rotating collapsing core, the
accretion disk, the protostar, the disk wind that gradually opens up
the outflow cavity, from different sets of initial conditions.  The
model grid covers a parameter space with $\mc=10-480~M_\odot$ and
$\scl=0.1-3~\gcm$, which is consistent with the observed environments
of massive star formation, and by sampling at different protostellar
masses, $\ms$, there are in total 432 different physical models in the
current model grid.  SEDs are generated via Monte-Carlo radiative
transfer simulation at 20 inclinations between an edge-on view and a
face-on view for each of these models, making a total of 8640 SEDs in
the model grid.  These model SEDs, also allowing for foreground
extinction, are used to fit the observed SED via $\chisq$
minimization.

In such a model grid, the properties and evolutions of the protostar
and its surrounding structures are more physically connected, which
reduces the dimensionality of the parameter spaces and the total
number of models.  It also helps to rule out possible fitting results
that are physically unrealistic or that are not internally
self-consistent.  Therefore, this model grid serves not only as a
fitting tool to estimate properties of massive protostars from
observed SEDs, but also as a test of core accretion theory.  Its use
tells us whether or not the observed SEDs of various massive
protostars can be explained by the core accretion theory, with
different initial conditions and evolutionary stages.

We studied how the parameters $\mc$, $\scl$, $\ms$, inclination
$\inc$, and foreground extinction $\av$ affect the various features of
the SEDs, especially the peak wavelength, the $20-40~\mu$m slope, the
$160-500~\mu$m slope, and the bolometric temperatures. All these
features show clear dependencies on the evolutionary stages.  
Among these features, the peak
wavelength of the SED and the $160-500~\mu$m slope are not so
sensitive to the inclination or possible foreground extinction, except
at an inclination close to face-on, while the $20-40~\mu$m slope or
the bolometric temperature are highly sensitive to the inclination.
The environmental mass surface density, $\scl$, also strongly affects
the $20-40~\mu$m slope, while the other features are only weakly
dependent on $\scl$.  We found that the degree of the flashlight
effect (the difference between the inferred luminosity from the SED
and the true total luminosity) is almost only dependent on the viewing
inclination and the opening angle of the outflow cavity.  With outflow
cavities with typical opening angles, the inferred luminosity can be
higher or lower than the true total luminosity by a factor of a few
from a low inclination to a high inclination.  However, with a
foreground extinction, the inferred bolometric luminosity almost
always underestimates the true luminosity by a factor almost solely
dependent on the outflow cavity opening angle.

We used the massive protostar G35.20-0.74 as an example of SED fitting
with our model grid (see also De Buizer et al. 2017 for a less
detailed application to eight sources, including G35.20-0.74).  The
fitting program not only provides information of a few best-fitted
models, but also shows the distribution of the fitted models in the
parameter space to understand constraints and degeneracies.  The
fitting yields a protostellar mass $\ms\approx10-20~M_\odot$, a total
luminosity of $(4-8)\times 10^4~L_\odot$, an accretion rate of a few
$\times 10^{-4}~M_\odot~\mathrm{yr}^{-1}$, and a half opening angle of
the outflow cavity of about $20^\circ$, which are consistent with
those estimated from other observations. The fitting also yields an
initial core mass of $\gtrsim 100~M_\odot$, while $\scl$ is not well
constrained.  There are certain degeneracies caused by combined
effects of $\scl$ and $\inc$.  Further breaking these degeneracies
will require additional observational constraints, such as using
predictions of image intensity profiles (e.g., \citealt[]{Zhang13}) or
radio continuum emission that traces ionized gas (e.g.,
\citealt[]{Tanaka16}).  Compared with the widely used
\citet[]{Robitaille06,Robitaille07} model grid (results presented by
\citealt[]{Debuizer17}), our model grid yields slightly lower
protostellar mass, similar total luminosities, but much higher
accretion rates (accretion rates of a few $\times
10^{-7}~M_\odot~\mathrm{yr}^{-1}$ are estimated using \citet[]{Robitaille06,Robitaille07} model).  
We believe that these differences are due, at least in
part, to there being a wider choice of free parameters in the
\citet[]{Robitaille06,Robitaille07} grid, which can lead to models that we consider less
physically realistic, i.e., high mass infall rates in the core
envelope but small disk accretion rates.  Our model grid, on the other
hand, is designed to include the different components more
consistently with fewer free parameters, to yield results that are
more physically realistic.

Future papers in this series will present the multi-wavelength imaging
data, which, as mentioned, may be helpful to break model
degeneracies. Extension to lower core masses is also planned (see
\citealt[]{ZT15} for some preliminary examples). 
These physical models, i.e., for the time evolution of density and
temperature, are also the necessary boundary conditions for
astrochemical computations and eventual line radiative transfer
simulations to predict molecular line emission properties of the
protostars.

\acknowledgments JCT acknowledges NSF grant AST-1411527. We
acknowledge the UF HPC and RIKEN HOKUSAI GreatWave for supporting
computational resources.

\end{document}